\documentclass[letterpaper, 12pt]{article} 
\usepackage[top=1in, bottom=1in, left=1in, right=1in]{geometry}

\usepackage{amsmath,amssymb,amsthm}
\usepackage{epsfig}
\usepackage{arydshln}
\usepackage{mathscinet}
\usepackage{subcaption}
\usepackage{color}

\newcommand{\bfm}[1]{\mbox{\boldmath ${#1}$}}
\newcommand{\nonum}{\nonumber \\}
\newcommand\eq[1] {(\ref{#1})} 
\newcommand{\beqa}{\begin{eqnarray}}
\newcommand{\eeqa}[1]{\label{#1}\end{eqnarray}}
\newcommand{\beq}{\begin{equation}}
\newcommand{\eeq}[1]{\label{#1}\end{equation}}
\newcommand{\Grad}{\nabla}
\newcommand{\Div}{\nabla \cdot}

\newcommand{\Real}{\mathop{\rm Re}\nolimits}
    \newcommand{\Imag}{\mathop{\rm Im}\nolimits}

\newcommand{\Md}{\partial}

\newcommand{\Ga}{\alpha}
\newcommand{\Gb}{\beta}
\newcommand{\Gd}{\delta}
\newcommand{\Ge}{\epsilon}
\newcommand{\Gve}{\varepsilon}

\newcommand{\Gg}{\gamma}
\newcommand{\Gc}{\chi}

\newcommand{\Gk}{\kappa}

\newcommand{\Gl}{\lambda}
\newcommand{\Gn}{\eta}
\newcommand{\Gm}{\mu}

\newcommand{\Gt}{\theta}

\newcommand{\Go}{\omega}

\newcommand{\GO}{\Omega}

\newcommand{\BGs}{\bfm\sigma}

\newcommand{\BGG}{\bfm\Gamma}

\newcommand{\BGY}{\bfm\Psi}

\newcommand{\CE}{{\cal E}}

\newcommand{\CH}{{\cal H}}

\newcommand{\CJ}{{\cal J}}

\newcommand{\CP}{{\cal P}}

\newcommand{\CS}{{\cal S}}
\newcommand{\CT}{{\cal T}}
\newcommand{\CU}{{\cal U}}

\newcommand{\bpm}{\begin{pmatrix}}
\newcommand{\epm}{\end{pmatrix}}
\newcommand\fig[1] {{\rm Figure}~\ref{fig:#1}}

\newcommand\labfig[1] {\label{fig:#1}}
\newcommand\sect[1] {\ref{sect:#1}}
\newcommand\labsect[1] {\label{sect:#1}}
\usepackage{url}
\usepackage{hyperref}

\def\Be{{\bf e}}

\def\Bj{{\bf j}}
\def\Bk{{\bf k}}

\def\Bn{{\bf n}}

\def\Bs{{\bf s}}

\def\Bu{{\bf u}}
\def\Bv{{\bf v}}

\def\Bx{{\bf x}}

\def\BA{{\bf A}}
\def\BB{{\bf B}}

\def\BE{{\bf E}}

\def\BG{{\bf G}}

\def\BI{{\bf I}}
\def\BJ{{\bf J}}

\def\BL{{\bf L}}
\def\BM{{\bf M}}

\def\BP{{\bf P}}
\def\BQ{{\bf Q}}

\usepackage{graphics}
\usepackage{graphicx}
\usepackage{amssymb}
\usepackage{float}

\title{{\bf An extremal problem arising in the dynamics of two-phase materials that directly reveals information about the internal geometry}}
\author{Ornella Mattei$^1$, Graeme W. Milton$^2$, and Mihai Putinar$^3$}
\date{\small{$^1$ Department of Mathematics, San Francisco State University, CA 94132, USA,\\
		$^2$Department of Mathematics, University of Utah, Salt Lake City, UT 84112, USA, \\
		$^3$Department of Mathematics, University of California at Santa Barbara, CA 93106, USA, 
		and School of Mathematics, Statistics and Physics, Newcastle University, NE1 7RU Newcastle upon Tyne, UK.
		\\Emails: mattei@sfsu.edu, milton@math.utah.edu, mputinar@math.ucsb.edu, mihai.putinar@ncl.ac.uk}}
\begin{document}
	\maketitle
	\vspace{2ex}
\begin{abstract}
  \noindent
  In two phase materials, each phase having a non-local response in time, it has been found
  that for some driving fields the response somehow untangles at specific times, and
  allows one to directly infer useful information about the geometry of the material, such as
  the volume fractions of the phases. Motivated by this, and to obtain an algorithm for designing
  appropriate driving fields, we find approximate, measure independent, linear relations between the
  values that Markov functions take
  at a given set of possibly complex points, not belonging to the interval [-1,1] where the measure
  is supported.
  The problem is reduced to simply one of polynomial approximation of a given
  function on the interval [-1,1] and to simplify the analysis Chebyshev approximation is used.
  This allows one to obtain explicit estimates of the error of the approximation, in terms
  of the number of points and the minimum distance of the points to the interval [-1,1].
  Assuming this minimum distance is bounded below by a number greater than 1/2, the error
  converges exponentially to zero as the number of points is increased. Approximate linear relations
  are also obtained that incorporate a set of moments of the measure. In the context of the
  motivating problem, the analysis also yields bounds on the response at any particular time
  for any  driving field, and allows one to estimate the response at a given frequency using an
  appropriately designed driving field that effectively is turned on only for a fixed interval of
  time. The approximation extends directly to Markov-type functions with a positive semidefinite
  operator valued measure, and this has applications to determining the shape of an inclusion
  in a body from boundary flux measurements at a specific time, when the time-dependent
  boundary potentials are suitably tailored. 

\end{abstract}


\section{Introduction}
\setcounter{equation}{0}
Many systems have responses that are nonlocal in time as this is naturally
a consequence of the fact that it takes time for subelements of the system
to respond. Usually, this leads one to either: (1) examine the response at each,
or one or more, frequencies as the convolution in time characterizing the
response of the system becomes a simple multiplication in the frequency
domain; or (2) examine the response to a delta function or Heaviside function
as this directly reveals the integral kernel characterizing the response.
In this paper we show that desired information about the system can be
directly obtained from selectively designed input signals that are neither
at constant frequency, nor delta or Heaviside functions.

Our initial motivation comes from the work \cite{Mattei:2016:BRL, Mattei:2016:BRV} where we derived microstructure-independent bounds on the viscoelastic response at a given time $t$ of two-phase periodic composites (in antiplane shear) with prescribed volume fractions $f_1$ and $f_2=1-f_1$ of the
phases and with an applied average stress or strain prescribed as a function of time. We found that the bounds were sometimes extremely tight at particular times $t=t_0$: see \protect{\fig{Hfig:Max_r0f1isotropy}}. This was quite a surprise because the response of each phase is nonlocal in time, yet somehow this response is untangled
at these particular times. Thus, the bounds could be used in an inverse
fashion to determine the volume fractions from measurements at time $t_0$.
Determining volume fractions
of phases is important in the oil industry, where one wants to know
the proportions occupied by oil and water in the rock,
to detecting breast cancer, to assessing the porosity of sea-ice
and other materials, and even to determining the volume of holes in
swiss cheese. While our bounds were very tight at specific times in some examples,
they were far from tight at all times in other examples: see \protect{\fig{Hfig:Max_r0f1isotropy_mod}}.

At that time it was totally unclear as to whether appropriate input signals
could produce the desired tight bounds at
a specific time and, if so, what algorithm should be followed to
design these input signals. The primary goal of this paper is to address
this problem in the case
where the input function has a finite number of frequencies. In particular,
for the example of \protect{\fig{Hfig:Max_r0f1isotropy_mod}}, our algorithm produces
the much tighter bounds of \protect{\fig{Hfig:visco_new.pdf}}.
It is an open question as to whether one can find smooth input signals, containing a continuum
of frequencies, such that the response $\Real[v(t_0)]$ of the material at a specific moment of time $t_0$ is totally measure independent,
while $\Real[v(t)]$ has a smooth
dependence on $t$, with $\Real[v(t)]\to 0$ when $t\to -\infty$. The example of
\protect{\fig{Hfig:Max_r0f1isotropy}} suggests that it may be possible.

We emphasize that our results are applicable not just to determining the volume fractions of the phases in a two-phase composite but also determining the volume and shape of an inclusion in a body from exterior boundary measurements. This is
shown in Sections \sect{Cald} and \sect{Gen}. It is a classical
and important inverse problem with a long history and many contributions:
see \cite{Calderon:1980:IBV, Kohn:1984:DCB, Kohn:1987:RVM, Sylvester:1987:GUT, Sylvester:1993:LAI, Kang:1997:ICP, Alessandrini:1998:ICP, Ikehata:1998:SEI, Bruhl:2003:DIT,
  Capdeboscq:2003:OAE, Ammari:2004:RSI, Ammari:2007:PMT,
  Kirsch:2011:SOH, Mueller:2012:LNI, Kolokolnikov:2015:RMS} and references therein.

A secondary goal of this paper is to solve an accompanying mathematical
approximation problem which we now outline, and which
is essential to achieve the primary goal.

\begin{figure}[t]
\centering
\includegraphics[width=0.7\textwidth]{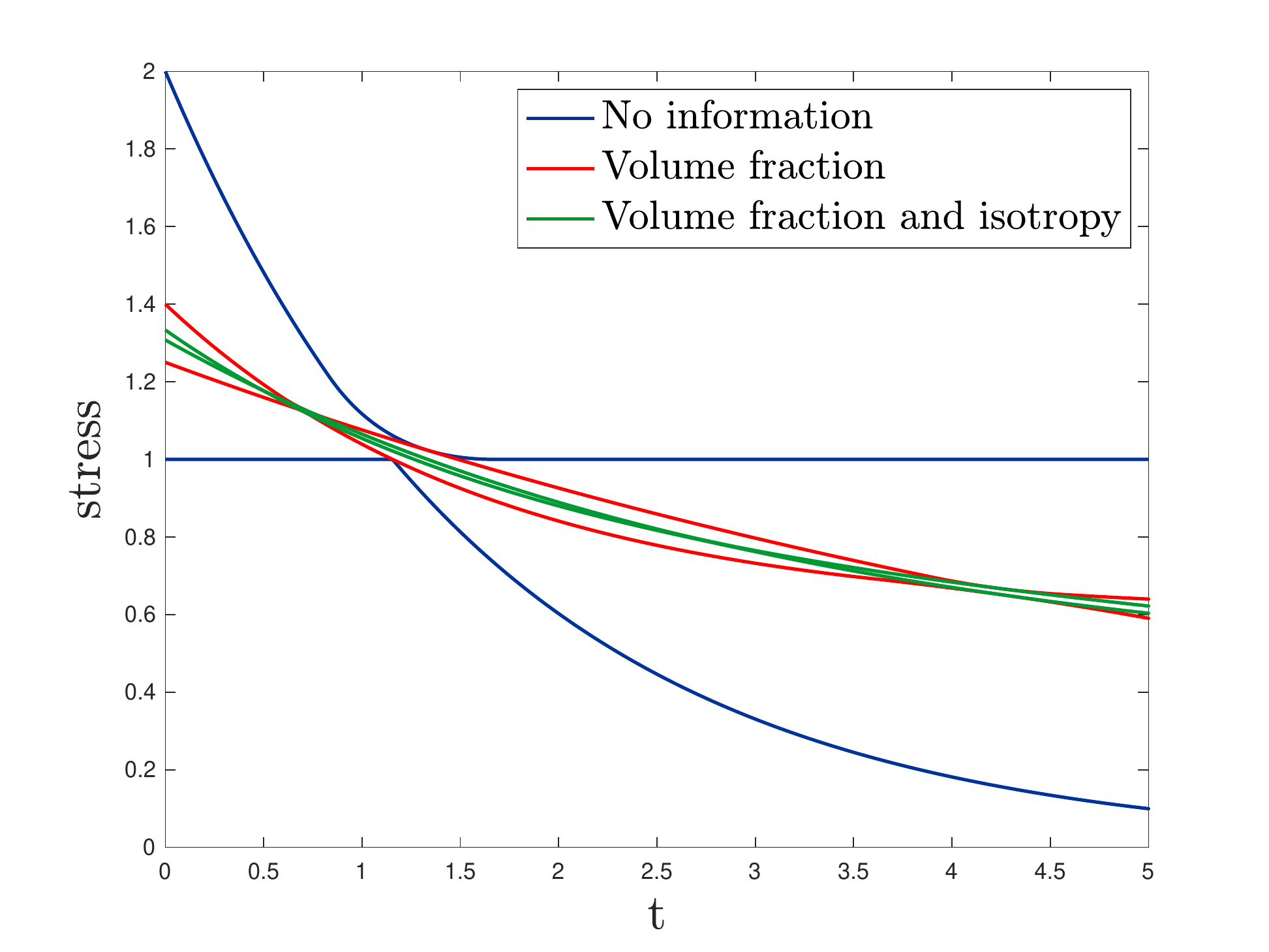}
\caption{{Comparison between the lower and upper bounds on the output average stress
  with an input
  applied average strain of $H(t)$, where $H(t)$ is the Heaviside function, 0 for $t<0$ and $1$ for $t\geq 0$.
  This is called a stress relaxation test. One phase is purely elastic ($G=6000$), while the other phase is viscoelastic and modeled by the Maxwell model ($G=12000$ and $\eta=20000$) (the results are normalized by the response of the elastic phase).
The following three cases are graphed: no information about the composite is given; the volume fraction of the components is known ($f_1=0.4$); and the composite is isotropic with given volume fractions. The bounds become tighter and tighter as more information on the composite structure is included,
so that if color is missing from the figure the outermost pair of bounds are those with no information, the middle pair include just the volume fraction, and
the innermost pair include both volume fraction and isotropy. Reproduced from Figure 6.2 in
\protect{\cite{Mattei:2016:BRV}}}.}
\labfig{Hfig:Max_r0f1isotropy}
\end{figure}

\begin{figure}[t]
\centering
\includegraphics[width=0.7\textwidth]{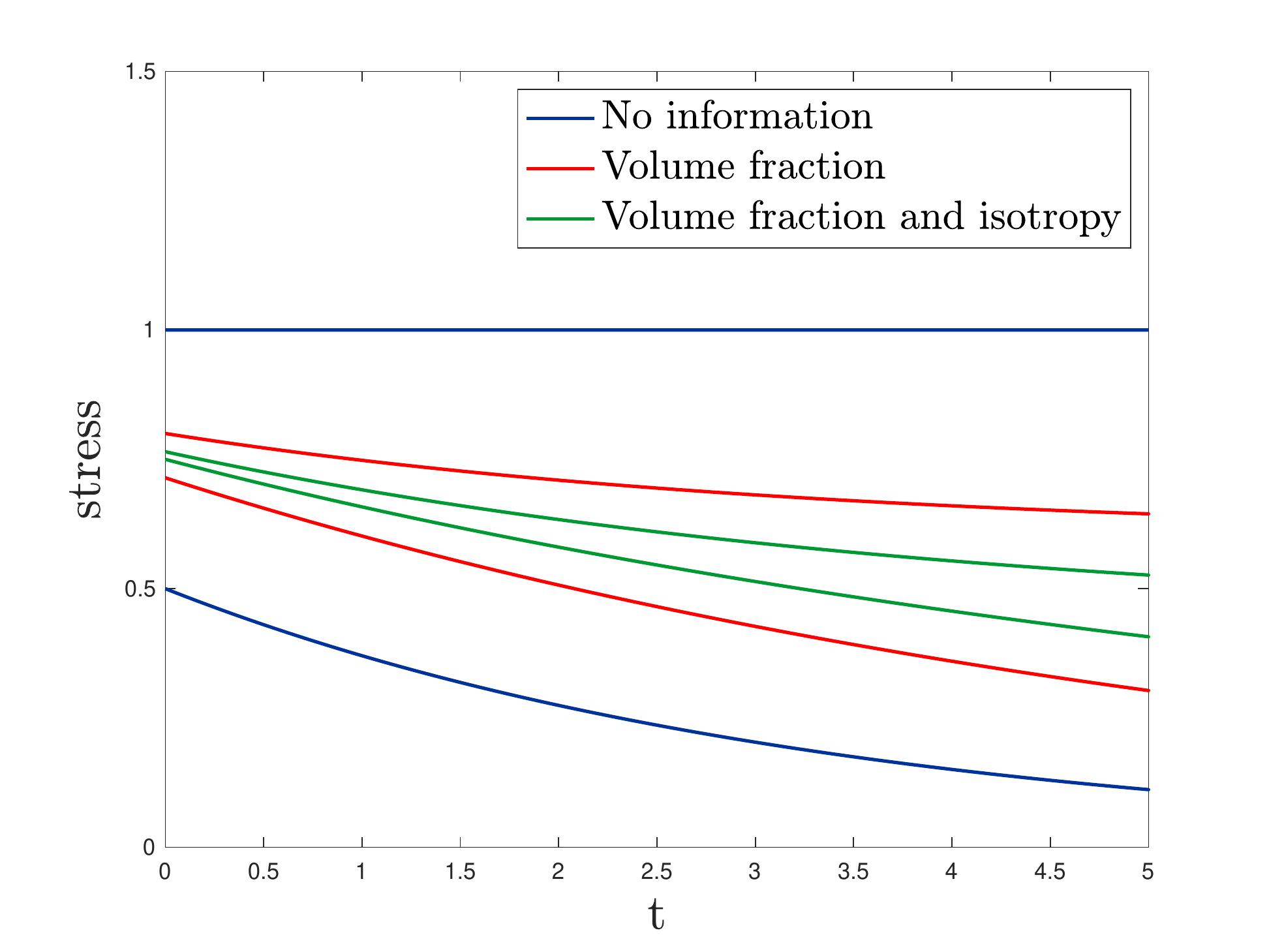}
\caption{{Comparison between the lower and upper bounds on
  the output stress relaxation in the ``badly-ordered case'', when
  the responses on the pure phases as a function of time do not cross with an input
  applied average strain of $H(t)$, where $H(t)$ is the Heaviside function. Here the purely elastic phase has shear modulus  $G=12000$, while the Maxwell parameters for the viscoelastic phase are $G=6000$ and $\eta=20000$ (again, the results are normalized by the response of the elastic phase).
  The three subcases are the same as for the previous figure.
However the bounds remain quite wide except near $t=0$. Reproduced from Figure 6.5 in
\protect{\cite{Mattei:2016:BRV}}}. The approach developed in this paper can yield tight bounds with a suitably designed
input function as shown in \protect{\fig{Hfig:visco_new.pdf}.}}
\labfig{Hfig:Max_r0f1isotropy_mod}
\end{figure}
\begin{figure}[t]
	\centering
	\includegraphics[width=0.7\textwidth]{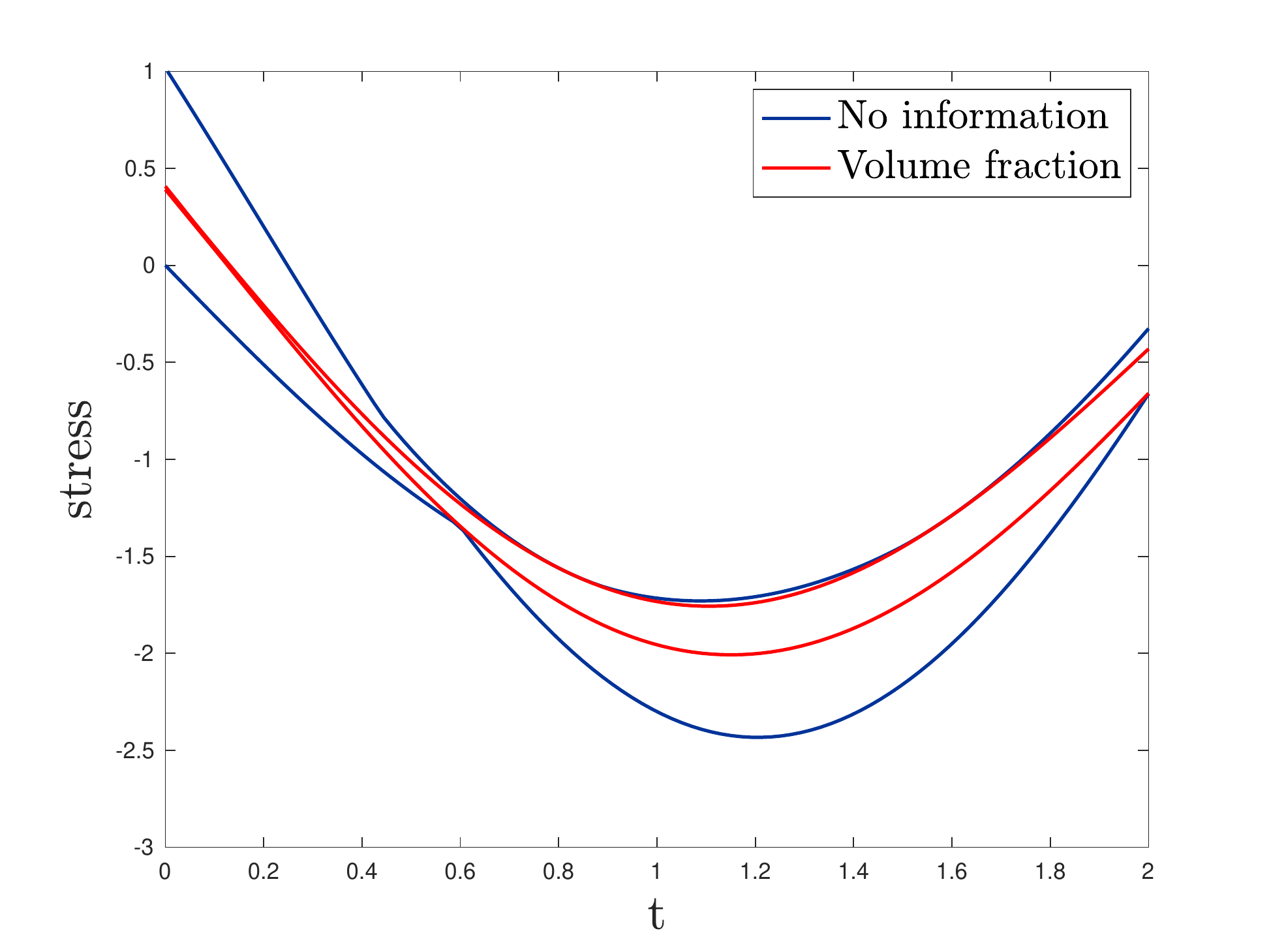}
	\caption{{Comparison between the lower and upper bounds on
			the output stress relaxation in the ``badly-ordered case'' ($G=12000$ for the elastic phase, and $G=6000$ and $\eta=20,000$ for the viscoelastic Maxwell phase), when
			the input function is chosen accordingly to equation \eqref{1.c}, which represents the main result of this paper. Specifically, equation \eqref{1.c} provides the amplitude of the applied field that gives extremely tight bounds at a chosen moment of time (here $t=0$) when the volume fraction is known. Indeed, the bounds incorporating the volume fraction (the innermost bounds, in red) take the value $0.4$ at $t=0$, which coincides exactly with the volume fraction of the viscoelastic phase. Here, the applied loading is the sum of three time-harmonic fields with frequencies $\omega=0.1,0.5,1.5$.}}
	\labfig{Hfig:visco_new.pdf}
\end{figure}

\section{{A fundamental mathematical question of
approximation applicable to Markov functions}}
\setcounter{equation}{0}

Here we formulate a problem, not just relevant to systems with a nonlocal
response, but potentially to other application where {\it Markov functions}
arise. Such functions are Cauchy transforms of positive measures with compact
support on the real line.  Some authors may suitably call them Herglotz functions, or Nevanlinna functions, or Stieltjes transforms.

Suppose $F_\mu(z)$ is a Markov function having the integral representation
\beq F_\mu(z)=\int_{-1}^1\frac{d\mu(\Gl)}{\Gl-z}, \eeq{0.1}
where the Borel measure $\mu$ is positive with unit mass:
\beq \int_{-1}^1 d\mu(\Gl)=1. \eeq{0.2}

Given $m$ (possibly complex) points $z_1,z_2,\ldots,z_m$ not belonging to the interval $[-1,1]$,  we are interested in finding 
complex constants $\Ga_1,\Ga_2,\ldots,\Ga_m$ such that
\beq \sum_{k=1}^m\Ga_k F_\mu(z_k)\approx 1 \eeq{0.3}
for all probability measures $\Gm$. 
Optimal bounds correlating the possible values of the $m$-tuple
$(F_\mu(z_1), F_\mu(z_2),\ldots, F_\mu(z_m))$ as $\mu$ varies over all probability measures are well known, as derived from the well charted analysis of the Nevanlinna-Pick interpolation problem \cite{Krein:1977:MMP}. Indeed, the nonlinear constraints among the values $F_\mu(z_1), F_\mu(z_2),\ldots, F_\mu(z_m)$,
and standard convexity theory provide optimal bounds on the range of the left hand side of \eq{0.3} for given constants $\Ga_1,\Ga_2,\ldots,\Ga_m$, see \cite{Krein:1977:MMP} for details.  But this is not our main concern.

We would rather like to choose $m$ points $z_1,z_2,\ldots,z_m$,
and find associated constants $\Ga_1,\Ga_2,\ldots,\Ga_m$, for every
prescribed integer $m$, having the property
\beq \sup_{\Gm}\left|\sum_{k=1}^m\Ga_k F_\mu(z_k)-1\right|\leq\Ge_m \eeq{0.3a}
for some bound $\Ge_m$ subject to $\Ge_m\to 0$ as $m\to\infty$. The geometry of the locus of these points is  obviously essential and it will be detailed in the sequel. The faster the convergence, the better.

 Since we deal with probability measures, condition \eq{0.3a} is equivalent to
 \beq \sup_{\Gl \in [-1,1]} \left| \sum_{k=1}^m \frac{\alpha_k}{\Gl-z_k} -1\right| \leq \Ge_m. \eeq{0.3b}
 And this is good news. We seek the minimal deviation from zero on the interval $[-1,1]$ of a rational function $R(z)$ satisfying $R(\infty) = 1$ and possessing simple poles at the points $z_1,\ldots,z_m$.
 Or equivalently, denoting $q(z) = (z-z_1) (z-z_2) \ldots (z-z_m)$ and $w(\Gl) = |q(\Gl)|^{-1}$ we aim at finding the minimal deviation from zero of a {\it monic polynomial} $p$ of degree $m$,
 with respect to the weighted norm $ \| p\  w \|_\infty = \sup_{\Gl \in [-1,1]} |p(\Gl)w(\Gl)|.$

 Both perspectives align to well-known classical studies in approximation theory. The first one is an extremal problem in rational approximation with prescribed poles,
 a subject going back at least to Walsh \cite{Walsh:1932:IAR}. A great deal of information in this respect was systematized in Walsh's book \cite{Walsh:1965:IAR}.
 The second approach is a genuine weighted Chebyshev approximation problem, and here we are on solid ground. First, note that the functions
 \beq w(\Gl), \Gl w(\Gl), \Gl^2 w(\Gl), \ldots \eeq{0.3c}
 form a Chebyshev system on the interval $[-1,1]$, that is, they are linearly independent and any linear combination of $w(\Gl), \Gl w(\Gl), \Gl^2 w(\Gl), \ldots, \Gl^m w(\Gl)$
 has at most $m$ zeros in $[-1,1]$. Even more, a stronger so-called Markov property of this system of functions holds. The classical Chebyshev approximation in the uniform norm theorem
 has an analog for such non-orthogonal bases \cite{Karlin:1966:TSA,Krein:1977:MMP}. To be more precise, there exists a unique monic polynomial $p$ of degree $m$
 minimizing the norm $\| p \ w \|_\infty$: this polynomial is characterized by the fact that $|p(\Gl)|$ attains its maximal value at $m+1$ points,
 and the sign of $p(\Gl)$ alternates there, see also \cite{Micchelli:1974:CCA}.
 In case $w(\Gl) =1$, the optimal polynomial is of course the normalized
 Chebyshev polynomial of the first kind: $p(\Gl) = \frac{T_m (\Gl)}{2^m}, \ T_m(\cos x) = \cos ( m x),\, m \geq 0.$ The constructive aspects of weighted
 Chebyshev approximation are rather involved, see for instance the early works of Werner \cite{Werner:1962:KET,Werner:1962:SDT,Werner:1962:TAB}.
 In the same vein, the asymptotics of the optimal bound of our minimization problem inherently involves potential theory or operator theory concepts.
 We cite for a comparison basis a few remarkable results of the same flavor \cite{Gonchar:1978:MTM,Prokhorov:2015:RAM,Baratchart:2001:BMA}.
 
 Without seeking sharp bounds and guided by the specific applications we aim at, we propose a compromise and relaxation of our extremal problem:
 \beq \inf_p \| p \ w \|_\infty \leq \| w \|_\infty \inf_p \| p \|_\infty. \eeq{0.3d}
 At this point we can invoke Chebyshev original theorem and his polynomial $T_m$, obtaining in this way the benefit, very useful for applications, of computing in closed form the residues 
 $\alpha_k$. Details and some ramifications will be given in the Section \sect{Main}.
 \section{{Relevance of the approximation problem
     to systems with a non-local time response and the viscoelasticity problem in particular.}}
\setcounter{equation}{0}

Without going into the specific details, as these will be provided later in Section 7, 8 and 9, in many linear systems with an input function $u(t)$ varying with time $t$, of the form
\beq u(t)=\sum_{k=1}^m\Gb_k e^{-i\Go_k(t-t_0)}, \eeq{0.4}
where the $\Go_k$ are a set of (possibly complex) frequencies, and $t_0$ is a given time,
the output function $v(t)$ takes the form
\beq v(t)=\sum_{k=1}^m\Ga_k a_0 F_\mu(z(\omega_k))e^{-i\Go_k(t-t_0)}, \eeq{0.5}
in which the function $F_\mu(z)$ is given by \eq{0.1},
\beq \Ga_k=\Gb_k c(\Go_k), \eeq{0.5a}
and the functions  $z(\omega)$ and $c(\omega)$ depend on $\Go$ in some known way: $z=z(\omega)$ and $c=c(\Go)$. The real constant $a_0>0$ and the unknown measure $d\mu$ depend on the system. In our viscoelasticity study \cite{Mattei:2016:BRV}
  the connection with Markov functions comes from the fact that the
effective shear modulus $G_*(\Go)$, that relates the average stress to the average strain at frequency $\Go$, as a function of the shear moduli $G_1(\Go)$ and
$G_2(\Go)$ of the two phases, has the property that
$[(G_*/G_1)-1]/(2f_1)$, in which $f_1$ is the volume fraction of phase 1,
is a Markov function of $z=(G_1+G_2)/(G_2-G_1)$ taking the form \eq{0.2}
\cite{Bergman:1978:DCC, Milton:1981:BCP, Golden:1983:BEP}.

Henceforth, we adopt the notational simplification
$$ f(z) = F_\mu(z).$$
Thus, at time $t=t_0$, the output function is
\beq v(t_0)=a_0\sum_{k=1}^m\Ga_kf(z_k) \quad\text{with}\quad z_k=z(\omega_k), \eeq{0.6}
and we seek an input signal so that the output $v(t_0)$ is almost system independent with $v(t_0)\approx a_0$. So, by measuring $v(t_0)$ we can determine the
system parameter $a_0$. In the viscoelastic problem that we studied \cite{Mattei:2016:BRL, Mattei:2016:BRV}, $a_0$ is the volume fraction $f_1$ (see also \cite{Bergman:1978:DCC}) and it is useful to be able to determine this from indirect measurements. 
Typically, one may assume the frequencies $\omega_k$
have a positive imaginary part so that the input signal $u(t)$ is essentially zero in the distant past. In \eq{0.4} one could just take a signal with $m-1$ frequencies
$\Go_k$, $k=1,2,\ldots, m-1$. Then, we have
\beq v(t_0)+\Ga_{m}a_0f(z(\omega_m))=a_0\sum_{k=1}^m\Ga_kf(z_k) \approx a_0 \quad\text{with}\quad z_k=z(\omega_k). \eeq{0.6a}
Then, if $a_0$ is known, a measurement of $v(t_0)$ will allow us to estimate the output $a_0f(z(\omega_m))e^{-i\Go_m(t-t_0)}$ at a desired (possibly real) 
frequency $\omega_m$ given the time harmonic input $e^{-i\Go_m(t-t_0)}$. 

It is often the situation, such as in the viscoelastic problem, that only the real part of $v(t)$ has a direct physical significance and, hence, one might want to find constants $\Ga_k$ such that, say,
\begin{align} 2\Real[v(t_0)]&=a_0\left(\sum_{k=1}^m\Ga_kf(z_k)+\sum_{k=1}^m\overline{\Ga_k}\overline{f(z_k)}\right)
\nonumber\\&=a_0\left(\sum_{k=1}^m\Ga_kf(z_k)+\sum_{k=1}^m\overline{\Ga_k}f(\overline{z_k})\right)\approx a_0,\label{0.7}
\end{align}
where the overline denotes complex conjugation. This, again, reduces to a problem of the form \eq{0.3} where, after renumbering, the complex values of $z_k$ come in pairs, $z_k$ and $z_{k+1}=\overline{z_k}$ and we may take $\alpha_{k+1}=\overline{\alpha_k}$ so that the left hand side of \eq{0.3} is real.

It may be the case that the first $n$ moments of the probability measure $d\Gm$ are known,
\beq M_\ell=\int_{-1}^1 \Gl^\ell d\mu(\Gl),\quad \ell=1,2,\ldots,n, \eeq{0.8}
in addition to the zeroth moment $M_0=1$
and that $m$ (possibly complex) points $z_1,z_2,\ldots,z_m$ not on the interval $[-1,1]$ are given. We then may seek complex constants 
$\Ga_1,\Ga_2,\ldots,\Ga_m$ and $\Gg_0,\Gg_1,\Gg_2,\ldots\Gg_n$, with say $\Gg_n=1$, such that 
\beq \sum_{k=1}^m\Ga_k f(z_k)\approx \sum_{\ell=0}^n\Gg_\ell M_\ell
\eeq{0.9}
for all probability measures $\Gm$ with the prescribed moments. We will treat this problem in Section 3. We can use these results to determine an 
approximate linear relation among the $n$ moments if $v(t_0)$ is measured. This may be used to estimate one moment if the rest are known. This can be
useful when the moments have an important physical significance: in the viscoelastic problem, for instance, $M_1$ depends only on the volume fraction $f_1$ if one assumes
that the composite has sufficient symmetry to ensure that its response remains invariant as the material is rotated \cite{Bergman:1978:DCC}. So, incorporating the moment $M_1$
and measuring the response at time $t_0$, then, allows us to obtain tighter bounds on $f_1$ as shown in \cite{Mattei:2016:BRL, Mattei:2016:BRV}. 

Mutatis mutandis, we may seek complex constants 
$\Ga_1,\Ga_2,\ldots,\Ga_m$ and $\Gg_1,\Gg_2,\ldots\Gg_n$ (each constant depending both on $m$ and $n$) such that
\beq \inf_\BA \left\|\sum_{k=1}^m\Ga_k[\BA-z_k\BI]^{-1}-\sum_{\ell=0}^n\Gg_\ell\BA^\ell\right\|\leq \Ge_m^{(n)},
\eeq{0.10}
where the infimum is over all self adjoint operators $\BA$ with spectrum in $[-1,1]$ and
for fixed $n$, $\Ge_m^{(n)}\to 0$ as $m\to\infty$. Our analysis extends easily to treat this problem too. The relevance is that 
in many linear systems with an input field $\Bu(t)$ varying with time $t$, of the form
\beq \Bu(t)=\sum_{k=1}^m\Gb_k e^{-i\Go_k(t-t_0)}\Bu_0, \eeq{0.11}
the output field $\Bv(t)$  takes the form
\beq \Bv(t)=\sum_{k=1}^m\Ga_k e^{-i\Go_k(t-t_0)}a_0[\BA-z(\omega_k)\BI]^{-1}\Bu_0\quad \text{with}\quad \Ga_k=\Gb_k c(\Go_k),
\eeq{0.12}
where the real constant $a_0$ and the self-adjoint operator $\BA$ characterize the response of the system, and the system parameters $z(\Go)$ and $c(\Go)$ depend on the frequency 
$\Go$ in some known way. Then, the bound \eq{0.10} implies
\beq \left|\Bv(t_0)-a_0\sum_{\ell=0}^n\Gg_\ell\BA^\ell\Bu_0\right|\leq a_0\Ge_m^{(n)}\left|\Bu_0\right|.
\eeq{0.13}

\section{The solution to the main approximation problem} 
\setcounter{equation}{0}
\labsect{Main}
%

The present section contains the main result which provides the theoretical foundation of our explorations. As explained in the introduction, we try to balance the computational accessibility and simplicity with the loss of sharp bounds. A few comments about the versatility of the following theorem are elaborated after its proof. 
\bigskip 

\noindent {\bf Theorem 1} \newline
{\it 
\smallskip
\noindent Letting 
\beq d(z_k)=\min_{\Gl\in[-1,1]}|\Gl-z_k| \eeq{1.a}
denote the distance from $z_k$ to the line segment $[-1,1]$, and assuming
\beq d_{\rm min}=\min_k d(z_k)>1/2, \eeq{1.b}
one can find complex constants $\Ga_1,\Ga_2,\ldots,\Ga_m$ each depending on $m$, such that \eq{0.3a} holds with $\Ge_m\to 0$ as $m\to\infty$. 
In particular, with
\beq \Ga_k=-\frac{T_m(z_k)}{2^{m-1}\prod_{j\ne k}(z_k-z_j)},
\eeq{1.c}
where $T_m(z)$ is the Chebyshev polynomial of the first kind, of degree $m$,
\eq{0.3a} holds with $\Ge_m=2/(2d_{\rm min})^m$ which tends to zero as $m\to\infty$. \newline
}
\bigskip

\noindent {\bf Proof}
\newline

\smallskip

\noindent We have
\beq \left|\sum_{k=1}^m\Ga_k f(z_k)-1\right|\leq \int_{-1}^1d\mu(\Gl)
\left|\sum_{k=1}^m\frac{\Ga_k}{\Gl-z_k}-1\right|. \eeq{1.3b}
Since the right hand side of \eq{1.3b} is linear in $d\Gm$ its maximum over all probability measures is achieved
when $\Gm(\Gl)$ is an extremal measure, namely the point mass
\beq \Gm(\Gl)=\Gd(\Gl-\lambda_0), \eeq{1.4}
where $\lambda_0$ is varied in $[-1,1]$ so as to get the maximum value of the right hand side of
\eq{1.3b}. In fact, for this extreme measure one has equality in \eq{1.3b}. Equivalently, we have
\beqa \inf_{\Ga}\sup_{\Gm}\left|\sum_{k=1}^m\Ga_k f(z_k)-1\right|& = & \inf_{\Ga}\sup_{\Gm}\int_{-1}^1d\mu(\Gl)
\left|\sum_{k=1}^m\frac{\Ga_k}{\Gl-z_k}-1\right| \nonum
& = & \inf_{\Ga}\sup_{\Gl_0\in[-1,1]}\left|\sum_{k=1}^m\frac{\Ga_k}{\Gl_0-z_k}-1\right|.
\eeqa{1.5}
Thus, we seek a set of constants $\Ga_1,\Ga_2,\ldots,\Ga_m$ (each dependent on $m$) and 
sequence $\Ge_m$ such that $\Ge_m\to 0$ as $m\to\infty$ and
\beq \left|\sum_{k=1}^m\frac{\Ga_k}{\Gl-z_k}-1\right| \leq \Ge_m\text{ for all }\lambda\in[-1,1].
\eeq{1.7}
More clearly, direct substitution of \eq{1.7} into \eq{1.3b} shows that \eq{0.3a} holds.

Now we may write
\beq \sum_{k=1}^m\frac{\Ga_k}{\Gl-z_k}=\frac{p(\Gl)}{q(\Gl)}=R(\Gl), \eeq{1.9}
where $q(\Gl)$ is the known monic polynomial
\beq q(\Gl)=\prod_{j=1}^m(\Gl-z_j) \eeq{1.10}
of degree $m$, and $p(\Gl)$ is a polynomial of degree at most $m-1$ that remains to be determined. The constants $\Ga_k$ can then be identified with the residues at the poles $\Gl=z_k$ of $R(\Gl)$:
\beq \Ga_k=\frac{p(z_k)}{\prod_{j\ne k}(z_k-z_j)}.
\eeq{1.10a}
Then, the problem becomes one of 
choosing $p(\Gl)$ such that
\beq  \sup_{\Gl\in[-1,1]}\left|\frac{p(\Gl)}{q(\Gl)}-1\right|=\sup_{\Gl\in[-1,1]}\frac{\left|p(\Gl)-q(\Gl)\right|}{|q(\Gl)|}
\eeq{1.11}
is close to zero. Clearly, the problem is now one of polynomial approximation of the monic polynomial $q(\Gl)$ of degree $m$ by the polynomial $p(\Gl)$. A natural choice is to take 
\beq p(\Gl)=q(\Gl)-T_m(\Gl)/2^{m-1}, \eeq{1.11a}
where $T_m(\Gl)/2^{m-1}$ is the Chebyshev polynomial $T_m(\Gl)$ of degree $m$, normalized to be monic. This choice minimizes the sup-norm of
$|p(\Gl)-q(\Gl)|$ over the interval $\lambda\in[-1,1]$ and 
\beq |p(\Gl)-q(\Gl)|=|T_m(\Gl)/2^{m-1}|\leq 1/2^{m-1} \eeq{1.12}
provides a bound on the numerator in \eq{1.11}. To bound the denominator, we have
\beq |q(\Gl)|= \prod_{k=1}^m|\Gl-z_k|\geq \prod_{k=1}^m d(z_k), \eeq{1.13}
where $d(z_k)$ is given by \eq{1.a}. Using \eq{1.b} and the bounds \eq{1.12} and \eq{1.13} we see that \eq{0.3a} is satisfied with
$\Ge^m=2/(2d_{\rm min})^m$. Finally, with $p(\Gl)$ given by \eq{1.11a} we see that the residues $\Ga_k$ at the poles $\lambda=z_k$ of $g(\Gl)$, given 
by \eq{1.10a} correspond to those given by \eq{1.c}.
\newline

\medskip
\noindent {\bf Remark 1}
\newline

\smallskip
\noindent
The use of Chebyshev polynomials is convenient as bounds on their sup-norm over the interval $[-1,1]$ are readily available. An alternative approach, also
accessible from the numerical/computational point of view, is to work with the
$L^2$ norm and find the polynomial $p(\Gl)$ of degree $m-1$ that approximates the given monic polynomial $q(\Gl)$ of degree $m$ in the precise sense that
\beq \int_{-1}^1 |(p(\Gl)-q(\Gl)|^2d\nu(\Gl), \quad\text{with}\quad d\nu(\Gl)=d\Gl/|q(\Gl)|^2
\eeq{1.15}
is minimized. Subsequently, one has to invoke Bernstein-Markov's inequality which bounds an $L^2$ norm by uniform norm.
This first step is a standard problem in the theory of orthogonal polynomials: one chooses $p(\Gl)-q(\Gl)$ to be the monic polynomial of degree $m$ that is
orthogonal to all polynomials of degree at most $m-1$ with respect to the measure $d\nu(\Gl)$. Separating the contribution of the denominator, by selecting $\nu$ to be the measure
$d\Gl/\sqrt{1-\Gl^2}$ we recover the Chebyshev polynomials we have advocated in the proof of the main result.
\newline 

\medskip
\noindent {\bf Remark 2}
\newline
\smallskip

\noindent Assumption \eq{1.b} is more than we really need. To underline the dependence on $n$ of all data we set
$$ q_n(z) = (z-z_1(n)) (z-z_2(n)) \ldots (z-z_n(n)), \ \ n \geq 1,$$
and
$$ w_n(z) = \frac{1}{|q_n(z)|}.$$
For the proof of Theorem 1 above we only need
$$ \limsup_n \| w_n \|_\infty^{1/n} < 2.$$
That is, there exists $r <2$, so that for large $n$, the inequality
$$ w_n(\Gl) \leq r^n, \ \ \Gl \in [-1,1],$$
holds true. 

By taking the natural logarithm, we are led to enforce the condition
$$ \limsup_n \sup_{\Gl \in [-1,1]} \frac{1}{n} \sum_{j=1}^n \ln \frac{1}{|\Gl-z_j(n)|} < \ln 2. $$
In other terms, an evenly distributed probability mass on the points $z_1(n), \ldots,z_n(n)$
should have its logarithmic potential asymptotically bounded from above by a prescribed constant, on the 
interval $[-1,1]$. Again, this turns out to be a rather typical problem of approximation theory, at least when restricting
the poles of $q_n$ to belong to some Jordan curve surrounding $[-1,1]$. A natural choice being an ellipse with foci at $\pm1$, see also \cite{Prokhorov:2015:RAM,Baratchart:2001:BMA}.
\newline

\medskip
\noindent {\bf Remark 3}
\newline

\smallskip
\noindent
We can gain more flexibility in the choice of the input signal if we replace $T_m(z_k)$ in the formula \eq{1.c} for the residues $\Ga_k$ with $(z_k-z_0)T_{m-1}(z_k)$,
where $z_0$ is a prescribed (possibly complex) zero of $p(\Gl)-q(\Gl)=(\lambda-z_0)T_{m-1}(\lambda)$. In particular,
we may choose $z_0$ to, say, minimize
\beq \max_{t\leq t_0} |v(t)|/|v(t_0)|\approx\max_{t\leq t_0}\left|\left[\sum_{k=1}^m\Ga_k f(z_k)e^{-i\Go_k(t-t_0)}\right]\right|
\eeq{1.20}
to help ensure that the output signal is not too wild. If we are only interested in $\Real[v(t)]$ so that the $z_k$ come in complex conjugate pairs,
then we may replace $T_m(z_k)$ in \eq{1.c} with $(z_k-z_0)(z_k-\overline{z_0})T_{m-2}(z_k)$, and choose $z_0$ to, say, minimize
\beq \max_{t\leq t_0} |\Real[v(t)]|/|\Real[v(t_0)]|\approx\max_{t\leq t_0}\left|\Real\left[\sum_{k=1}^m\Ga_k f(z_k)e^{-i\Go_k(t-t_0)}\right]\right|.
\eeq{1.21}
In the first case, note that the signal $u(t)$ \eqref{0.4} is linear in $z_0$ while in the second case it is linear in the real coefficients of the quadratic $(\lambda-z_0)(\lambda-\overline{z_0})$.
So in either case we have a linear space of possible signals (though $|z_0|$ should not be too large for the approximation to hold at time $t_0$). Also $\Ga_k\to 0$
as $z_0\to z_k$ so in this limit the frequency $\omega_k$ is absent from the input and output signals. More generally,
to help minimize \eq{1.20} or \eq{1.21} one might replace $T_m(z_k)$ with $s_M(z_k)T_{m-M}(z_k)$ where $s_M(\Gl)$ is a polynomial of fixed degree $M<m$.

\section{Incorporating moments of the measure}
\setcounter{equation}{0}
%
Here we assume that the first $n$ moments $M_1,M_2,\ldots,M_n$ of the probability measure $d\Gm$, given by \eq{0.1},
are known, in addition to $M_0=1$ and that $m$ (possibly complex) points $z_1,z_2,\ldots,z_m$ not on the interval $[-1,1]$ are given. We seek complex constants 
$\Ga_1,\Ga_2,\ldots,\Ga_m$ and $\Gg_1,\Gg_2,\ldots\Gg_n$, with say $\Gg_n=1$  such that 
\beq \left|\sum_{k=1}^m\Ga_k f(z_k)-\sum_{\ell=0}^n\Gg_\ell M_\ell\right|
\eeq{2.2}
is small for all probability measures $\Gm$ with the prescribed $n$ moments. The analysis proceeds as before, only now we introduce the polynomial
\beq r(\Gl)=\sum_{\ell=0}^n\Gg_\ell \Gl^\ell, \eeq{2.3}
and set $p(\Gl)$ and $q(\Gl)$ to be the polynomials defined by \eq{1.9} and \eq{1.10}.
The goal is now to choose polynomials $p(\Gl)$ and $r(\Gl)$ of degrees $m-1$ and $n$, respectively, such that $r(\Gl)$ is monic and 
\beq \sup_{\Gl\in[-1,1]}\left|\frac{p(\Gl)}{q(\Gl)}-r(\Gl)\right|=\sup_{\Gl\in[-1,1]}\frac{\left|p(\Gl)-q(\Gl)r(\Gl)\right|}{|q(\Gl)|}
\eeq{2.4}
is close to zero. We choose $p(\Gl)$ and $r(\Gl)$ such that 
\beq T_{m+n}(\Gl)/2^{m+n-1}=q(\Gl)r(\Gl)-p(\Gl). \eeq{2.5}
This is simply the Euclidean division of the normalized Chebyshev polynomial $T_{m+n}(\Gl)/2^{m+n-1}$ by $q(\Gl)$ with $r(\Gl)$ being identified as the quotient polynomial
and $-p(\Gl)$ being identified as the remainder polynomial. Then, assuming \eq{1.b} and using \eq{1.13}, we have 
\beq \sup_{\Gl\in[-1,1]}\left|\frac{p(\Gl)}{q(\Gl)}-r(\Gl)\right|\leq \Ge_m^{(n)},\quad \text{with} \quad \Ge_m^{(n)}=\frac{2}{2^n(2d_{\rm min})^m}
\eeq{2.6}
satisfying $\Ge_m^{(n)}\to 0$ as $m\to\infty$, with $n$ being fixed. With constants $\Ga_k$ given by \eq{1.10a} and
constants $\Gg_\ell$ being the coefficients of the polynomial $r(\Gl)$, as in \eq{2.3}, it follows that
 \beqa &~& \sup_{\Gm}\left|\sum_{k=1}^m\Ga_k f(z_k)-\sum_{\ell=0}^n\Gg_\ell M_\ell \right| \nonum
&~& \quad = \sup_{\Gm}\int_{-1}^1d\mu(\Gl)
\left|\sum_{k=1}^m\frac{\Ga_k}{\Gl-z_k}-\sum_{\ell=0}^n\Gg_\ell \Gl^\ell\right| \leq  \Ge_m^{(n)}.
\nonum &~&
\eeqa{2.7}


\medskip
\noindent {\bf Remark}
\newline

\smallskip
\noindent
The results exposed in the previous sections extend immediately to the resolvent case using the spectral representation
\beq \BA=\int_{\sigma(\BA)}\Gl d\BP_\Gl, \eeq{3.1}
where $\sigma(\BA)$ is the spectrum of $\BA$, assumed to be contained in the interval $[-1,1]$ and $d\BP_\Gl$ is an orthogonal projection valued 
measure satisfying
\beq \BI=\int_{\sigma(\BA)}d\BP_\Gl . \eeq{3.2}
Then, we have
\beqa &~& \inf_\BA \left\|\sum_{k=1}^m\Ga_k[\BA-z_k\BI]^{-1}-\sum_{\ell=0}^n\Gg_\ell\BA^\ell\right\|
\nonum &~& \quad\quad=
\sup_{d\BP_\Gl}\left\|\int_{\sigma(\BA)} \left[\sum_{k=1}^m \frac{\Ga_k}{\Gl-z_k}-\sum_{\ell=0}^n\Gg_\ell\Gl^\ell \right]  d\BP_\Gl \right\| \nonum
&~&\quad\quad
\leq \sup_{d\BP_\Gl}\left\|\int_{\sigma(\BA)} \left|\sum_{k=1}^m \frac{\Ga_k}{\Gl-z_k}-\sum_{\ell=0}^n\Gg_\ell\Gl^\ell \right|  d\BP_\Gl \right\|. 
\eeqa{3.3}
Choosing constants $\Ga_1,\Ga_2,\ldots,\Ga_m$ and $\Gg_1,\Gg_2,\ldots\Gg_n$, with $\Gg_n=1$, as in the previous section, the bound \eq{2.6} 
substituted in \eq{3.3} implies that the desired bound \eq{0.10} holds with $\Ge_m^{(n)}=2/[2^n(2d_{\rm min})^m]$.

\section{Bounds on the output function $v(t)$ at any time $t$}
\setcounter{equation}{0}
Supposing any constants $\Ga_1,\Ga_2,\ldots,\Ga_m$ are given, it is easy to get bounds on $v(t)$ given by \eq{0.5}
at any time $t$ that incorporate the $n$ known
moments $M_1,M_2,\ldots,M_n$. One introduces an angle $\Gt$ and Lagrange multipliers $\Gg_1, \Gg_2, \ldots, \Gg_n$ and takes the minimum value of
\beq \int_{-1}^1d\mu(\Gl)\Real\left[e^{i\Gt}\sum_{k=1}^m\frac{\Ga_k e^{-i\Go_k(t-t_0)}}{\Gl-z(\Go_k)}+\sum_{\ell=1}^n\Gg_\ell\Gl^\ell\right],
\eeq{4.1}
as $\Gm$ varies over all probability measures supported on $[-1,1]$ with unconstrained moments. The minimum will be achieved by the point 
masses $\Gm=\Gd(\Gl-\Gl_0)$, where $\Gl_0$ may take one or more values. Typically we will need to choose the Lagrange multipliers $\Gg_1, \Gg_2, \ldots, \Gg_n$ 
(that depend on $\Gt$) so that the minimum is achieved at $n$ values $\Gl_0=\Gl_0^{(\ell)}$, $\ell=1,2,\ldots, n$, and then adjust the measure to be distributed at these points 
\beq d\mu(\Gl)=\sum_{\ell=1}^n \mathrm{w}_\ell\Gd(\Gl-\Gl_0^{(\ell)}),
\eeq{4.2}
with the non-negative weights $\mathrm{w}_\ell$, that sum to $1$, chosen so that the moments take their desired values. Then with this measure we obtain the bound
\beq \Real[e^{i\Gt} v(t)]\geq a_0\sum_{\ell=1}^n \mathrm{w}_\ell\Real\left[e^{i\Gt}\sum_{k=1}^m\frac{\Ga_k e^{-i\Go_k(t-t_0)}}{\Gl_0^{(\ell)}-z(\Go_k)}\right].
\eeq{4.3}
By varying $\Gt$ from $0$ to $2\pi$ we obtain bounds that confine $v(t)$ to a convex region in the complex plane. Of course, if we are only interested in
bounding $\Real[v(t)]$, then it suffices to take $\Gt=0$ or $\pi$. 

\begin{figure}[h!]
		\begin{subfigure}[b]{\textwidth}
			\centering
			\includegraphics[width=\linewidth]{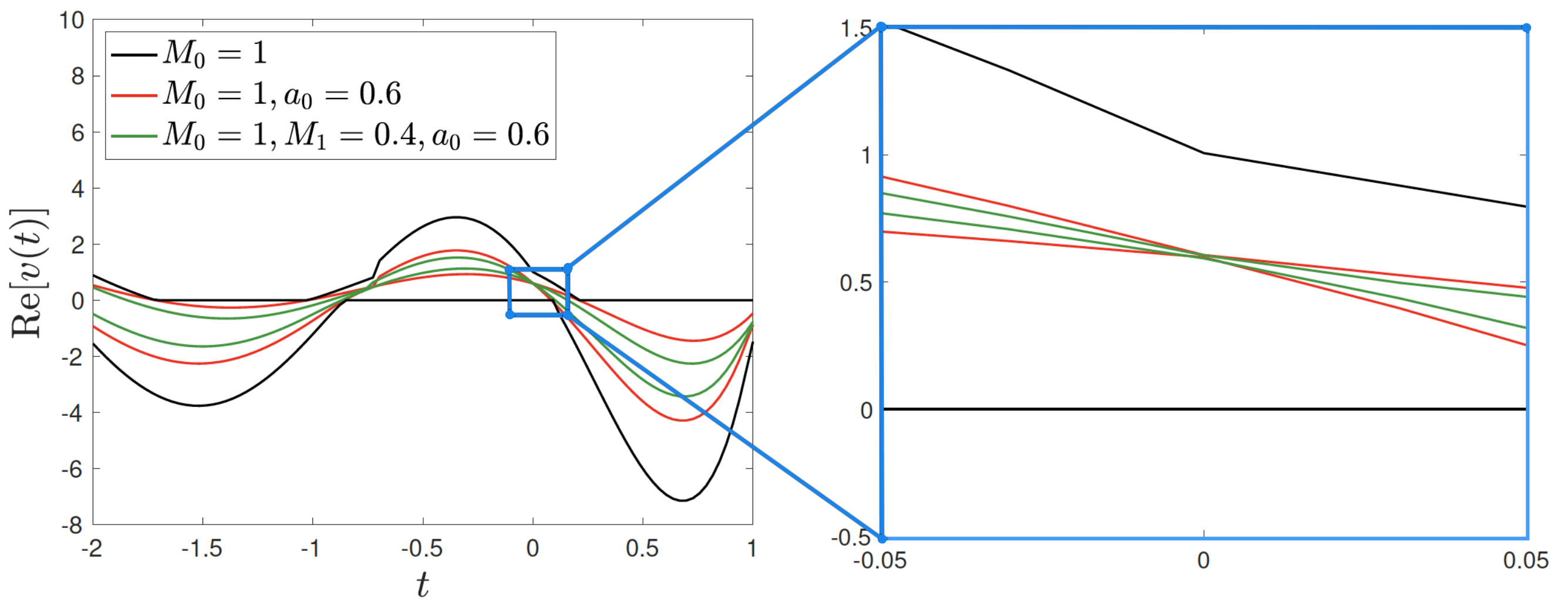}
			\caption{}
		\end{subfigure}
		
		\begin{subfigure}[b]{\textwidth}
			\centering
			\includegraphics[width=0.5\linewidth]{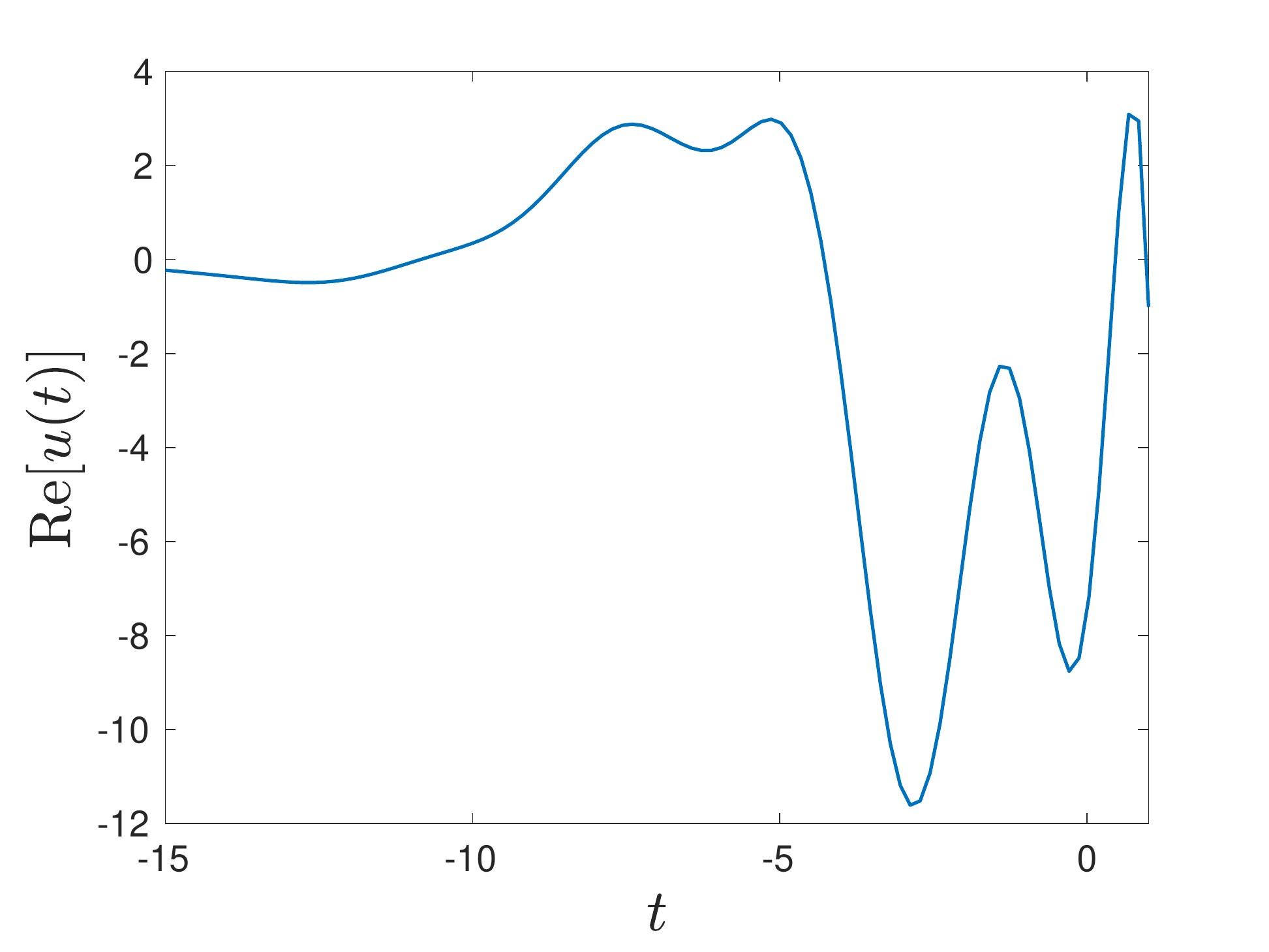}
			\caption{}
		\end{subfigure}
	\caption{(a) Bounds on the real part of the response of the system, $\mathrm{Re}[v(t)]$ \eqref{4.3}, when the system is such that $z(\omega)=2+i/\omega$ and the input signal $\mathrm{Re}[u(t)]$ is the one depicted in (b) (with $c(\omega)$=1). We choose the frequencies $\omega_k$ to be $[1+1i;0.5+0.3i;2+0.5i]$, and we select the coefficients $\alpha_k$ according to \eqref{1.c} so that the bounds are extremely tight at $t_0=0$, whereas the point masses $\lambda_0^{(\ell)}$ and the weights $\mathrm{w}_\ell$ are chosen for each moment of time $t$ such that the minimum value of \eqref{4.1} is attained while the moments of the measure take their desired values. Specifically, the bounds on $\mathrm{Re}[v(t)]$ are plotted for three different scenarios, as shown by the legend. }
\labfig{fig:bounds_Re_v}
\end{figure}

\begin{figure}[h!]
			\begin{subfigure}[b]{\textwidth}
				\centering
		\includegraphics[width=\linewidth]{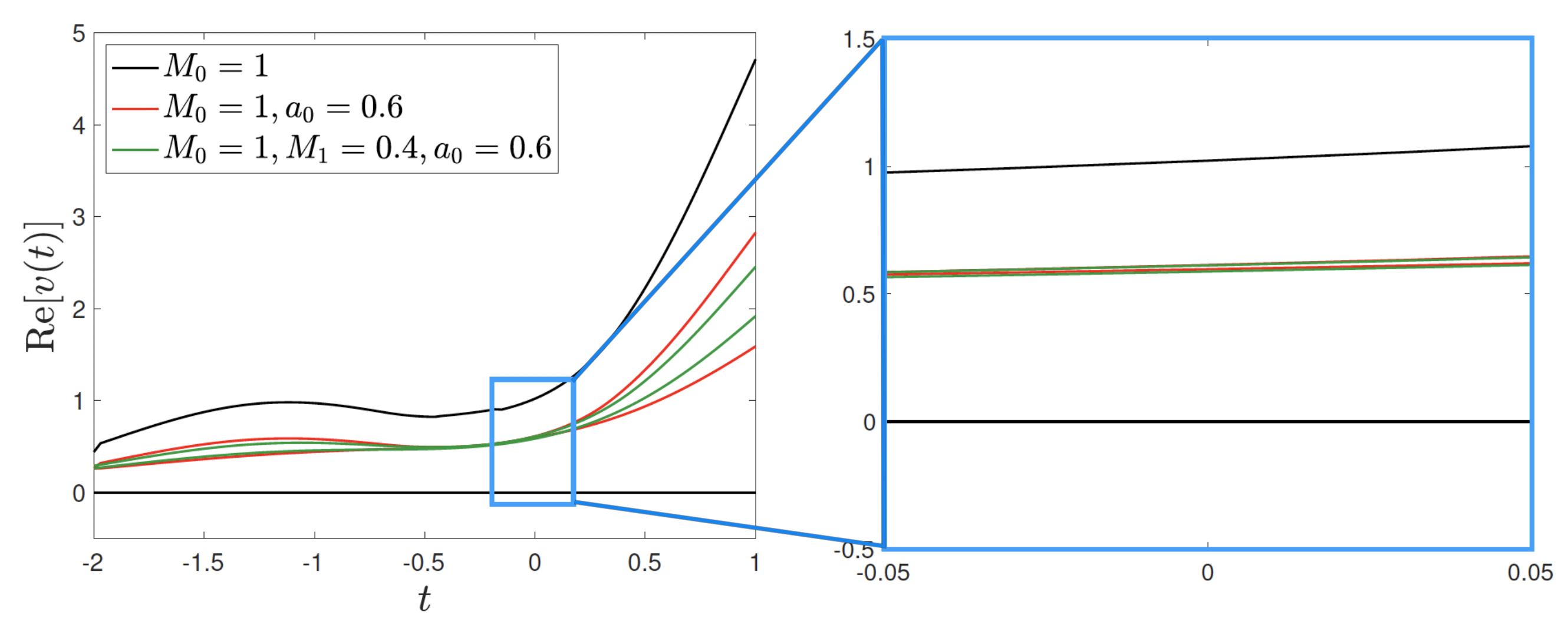}
		\caption{}
	\end{subfigure}
	
	\begin{subfigure}[b]{\textwidth}
		\centering
		\includegraphics[width=0.5\linewidth]{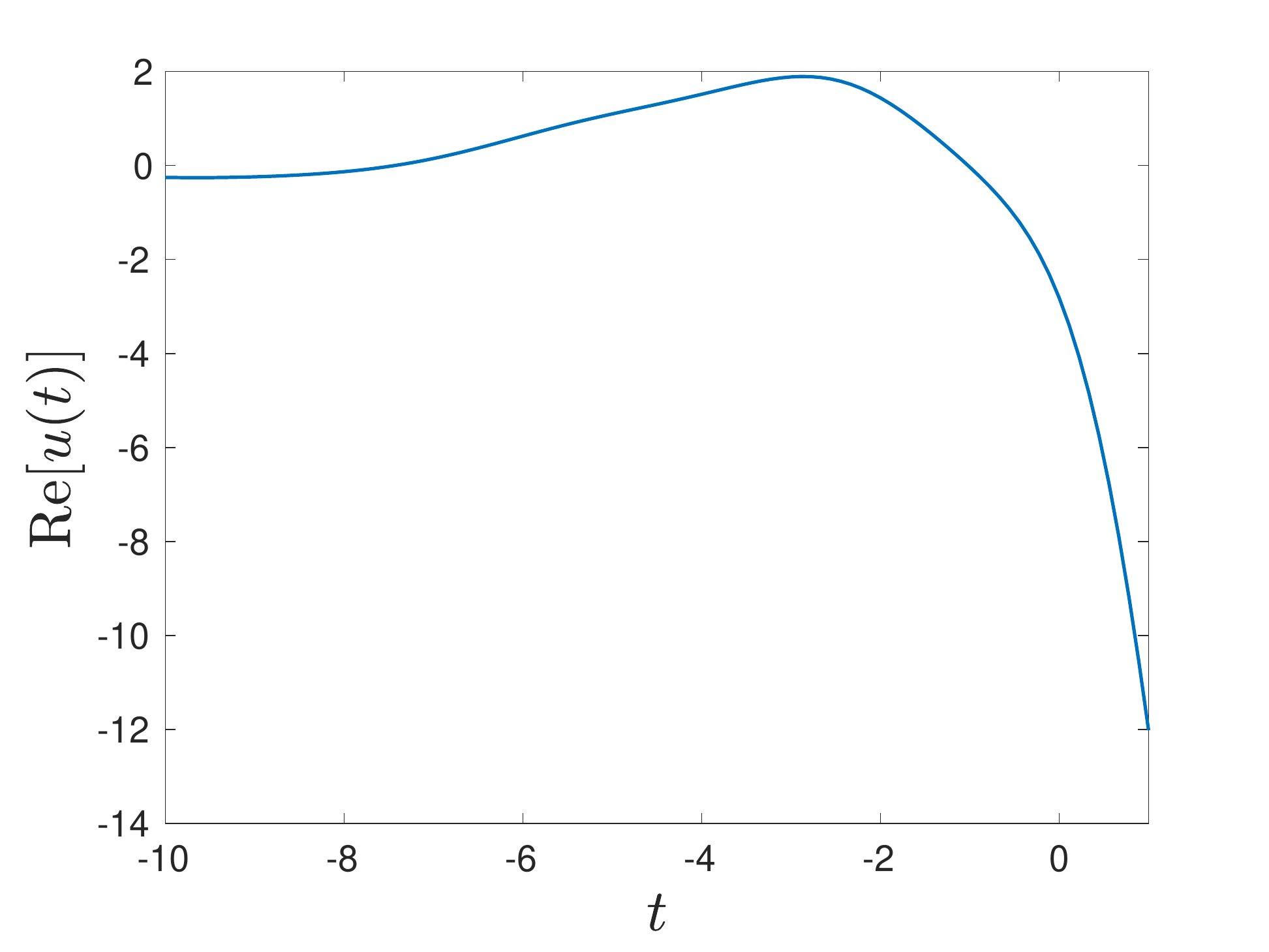}
		\caption{}
	\end{subfigure}
	\caption{(a) Bounds on the real part of the response of the system, $\mathrm{Re}[v(t)]$ \eqref{4.3}, when the system is such that $z(\omega)=2-2/\omega^2$ and the input signal $\mathrm{Re}[u(t)]$ is the one depicted in (b) (with $c(\omega)$=1). We choose the frequencies $\omega_k$ to be $[1+1i;0.5+0.3i;2+0.5i]$, like in the case depicted in \protect{\fig{fig:bounds_Re_v}}.}
	\labfig{fig:bounds_Re_v_model2}
      \end{figure}

\protect{\fig{fig:bounds_Re_v}} and \protect{\fig{fig:bounds_Re_v_model2}} depict the lower and upper bounds on $\mathrm{Re}[v(t)]$ for two systems ($z(\omega)=2+i/\omega$ in \protect{\fig{fig:bounds_Re_v}}, thus mimicking the low frequency dielectric response of a lossy dielectric material, and $z(\omega)=2-2/\omega^2$ in \protect{\fig{fig:bounds_Re_v_model2}}, thus mimicking the dielectric response of a plasma), when the coefficients $\alpha_k$ in \eqref{4.3} are chosen such that the bounds are extremely tight at $t_0=0$, according to \eqref{1.c}. For both systems, the bounds on $\mathrm{Re}[v(t)]$ are tighter the higher the amount of pieces of information on the system is incorporated. Notice that the bounds colored in black (the largest ones) correspond to the case where only the zeroth order moment $M_0$ of the measure is known but not the value of $a_0$: in such a case, as shown by the zoomed graph in the blue box, at $t=0$, the upper bound takes value 1 and the lower bound takes value 0, that are the smallest and the highest values $a_0$ can take. On the other hand, when $a_0$ is assigned, the value that the corresponding bounds take at $t=0$ is exactly $a_0=0.6$, as shown by the zoomed graph in the blue box. The graphs show clearly that, in order to estimate the system parameter $a_0$, one has just to measure the response of the system at a specific moment of time $t_0$ (if the applied field is carefully chosen).

      {These are the type of bounds used in \cite{Mattei:2016:BRV} to bound the temporal response of
        two-phase composites in
      antiplane elasticity. It is not yet clear whether those bounds can be derived from variational principles.
      In general, in the theory of composites,  variational methods have proven to be more powerful than analytic approaches.
      Variational methods produce tighter bounds that often easily extend to multiphase composites: see the
      books \cite{Cherkaev:2000:VMS, Torquato:2001:RHM, Milton:2002:TOC, Allaire:2002:SOH, Tartar:2009:GTH}
        and references therein. 
      For example, the variational approach gives tighter bounds on the complex permittivity at constant frequency
      of two-phase lossy composites \cite{Kern:2020:RCE}, than the bounds obtained by the
      analytic approach \cite{Milton:1981:BCP, Bergman:1982:RBC}.
      It also produces bounds on the
      complex effective bulk and shear moduli of viscoelastic composites \cite{Gibiansky:1993:EVM, Milton:1997:EVM}.
      An exception is bounds that correlate the
      complex effective dielectric constant at more than two frequencies \cite{Milton:1981:BTO}
      that have yet to be obtained by a systematic variational
      approach. Variational bounds in the time domain are available \cite{Carini:2015:VFL, Mattei:2017:BOP},
      but these are nonlocal in time.}

\section{Using an appropriate input signal to predict the response at a given frequency}
\setcounter{equation}{0}
Naturally, if one is interested in the response $v_0(t)$ at a given (possibly complex) frequency $\Go_0$, the easiest solution is to take
an input signal $u_0(t)$ at that frequency. However, it might not be easy to experimentally generate a signal at that frequency or it
might not be easy to measure the response at that frequency. The problem becomes: find complex constants $\Ga_1,\Ga_2,\ldots, \Ga_m$ such that 
\beq \sup_{\Gl \in [-1,1]} \left| \sum_{k=1}^m \frac{\alpha_k}{\Gl-z_k} -\frac{1}{\Gl-z_0}\right|
\leq \Ge_m,
\eeq{a.1}
with $z_k=z(\omega_k)$, $k=0,1,\dots,m$. Defining the polynomials $p(\Gl)$ and $q(\Gl)$ as in \eq{1.9} and \eq{1.10} one needs to find $p(\Gl)$ of degree $m-1$ such
\beq \sup_{\Gl \in [-1,1]} \left|\frac{(\Gl-z_0)p(\Gl)-q(\Gl)}{(\Gl-z_0)q(\Gl)}\right|\leq \Ge_m. \eeq{a.2}
Proceeding as before we choose
\beq (\Gl-z_0)p(\Gl)=q(\Gl)-b_mT_{m-1}(\Gl)\quad \text{with}\,\, b_m=q(z_0)/T_{m-1}(z_0), \eeq{a.3}
where $b_m$ has been chosen so that the polynomial $q(\Gl)-b_mT_{m-1}(\Gl)$ has a factor of $(\Gl-z_0)$. Then the residues of $R(\Gl)=p(\Gl)/q(\Gl)$ are given by
\beq \Ga_k=-b_m\frac{T_{m-1}(z_k)}{(z_k-z_0)\prod_{j\ne k}(z_k-z_j)}=-\frac{T_{m-1}(z_k)\prod_{j\ne 0}(z_0-z_j)}{T_{m-1}(z_0)(z_k-z_0)\prod_{j\ne 0,k}(z_k-z_j)},
\eeq{a.3a}
and
\beq \sup_{\Gl \in [-1,1]} |(\Gl-z_0)p(\Gl)-q(\Gl)|=\sup_{\Gl \in [-1,1]}|b_mT_{m-1}(\Gl)|=|b_m|, \eeq{a.4}
so that \eq{a.1} holds with
\beq \Ge_m=\frac{|b_m|}{d_0 \inf_{\lambda \in [-1,1]}| q(\lambda)|},
\eeq{a.5}
where $d_0$ denotes the distance from $z_0$ to the interval [-1,1].  Joukowski's map yields:
\beq  z_0 = \frac{1}{2}\left( \zeta_0 + \frac{1}{\zeta_0}\right),\,\, \text{with}\,\, R = |\zeta_0|>1, \eeq{a.5a}
whence
\beq T_{m-1}(z_0) = \frac{1}{2}\left(\zeta_0^{m-1} + \frac{1}{\zeta_0^{m-1}}\right). \eeq{a.5b}
Moreover, since $\zeta_0$ runs over a circle of radius $R$, we have
\beq d_0 = \inf_{\lambda \in [-1,1]} \frac{ | \zeta_0^2 - 2\zeta_0 \lambda+1|}{2} \geq \frac{(R-1)^2}{2R}, \eeq{a.5c}
and
\beq  |T_{m-1}(z_0)| \geq \frac{1}{2} (R^{m-1} - R^{1-m}), \ \ m \geq 2, \eeq{a.5d}
implying
\beq |b_m| \leq \frac{2 |q(z_0)|}{R^{m-1} - R^{1-m}}. \eeq{a.5e}
All in all, the relevant bound $\Ge_m$ satisfies
\beq |\Ge_m| \leq \frac{4R}{(R-1)^2} \frac{1}{R^{m-1}-R^{1-m}} \sup_{\lambda \in [-1,1]} \frac{|q(z_0)|}{|q(\lambda)|}. \eeq{a.5f}
We obtain an exponential decay $\Ge_m \rightarrow 0$ as $m \rightarrow \infty$ provided the geometry of the loci
$z_1,z_2,\ldots,z_m$ is subject to the following condition: for a positive constant $r<R$, each $z_j\in H(r)=H_1(r)\cup H_2(r)\cup H_3(r)$ where 
\beqa 
H_1(r) & = & \left\{z:\left|\frac{z - z_0}{z+1}\right| \leq r, \ \ \Real z \leq -1\right\}, \nonum
H_2(r) & = & \left\{z :\left|\frac{z-z_0}{\Imag z} \right|\leq r, \ \ \Real z \in [-1,1] \right\},\nonum
H_3(r) & = & \left\{z :\left|\frac{z - z_0}{z-1}\right| \leq r, \ \ \Real z \geq 1\right\}.
\eeqa{a.5g}
In other words, all of the $z_j$ must be close to $z_0$ in the precise sense that $z_j\in H(r)$.
Note that, as shown in \protect{\fig{fig:loci}}a, in case $r <1$, $H_1$ and $H_3$ are sectors of disks, while $H_2$ is a portion of an ellipse. 
For $r \in (1,R)$ these regions are complements of disks/ellipse, containing the point $z_0$, as shown in \protect{\fig{fig:loci}}c. Some of these regions can be empty, depending on the position of $z_0$.
\begin{figure}[h!]
	\begin{subfigure}[b]{0.5\textwidth}
		\centering
		\includegraphics[scale=.5]{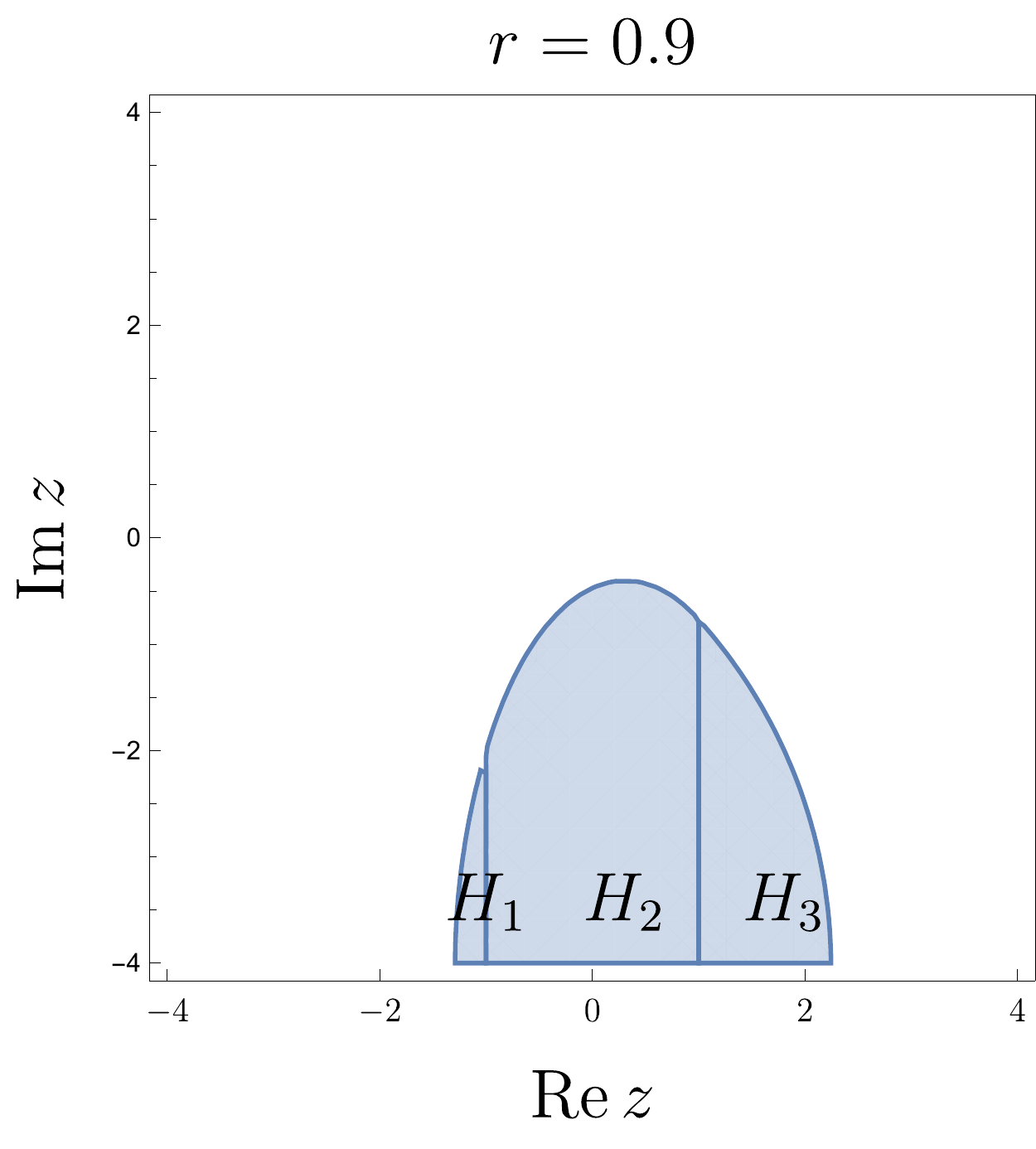}
		\caption{}
	\end{subfigure}
\begin{subfigure}[b]{0.5\textwidth}
	\centering
	\includegraphics[scale=.5]{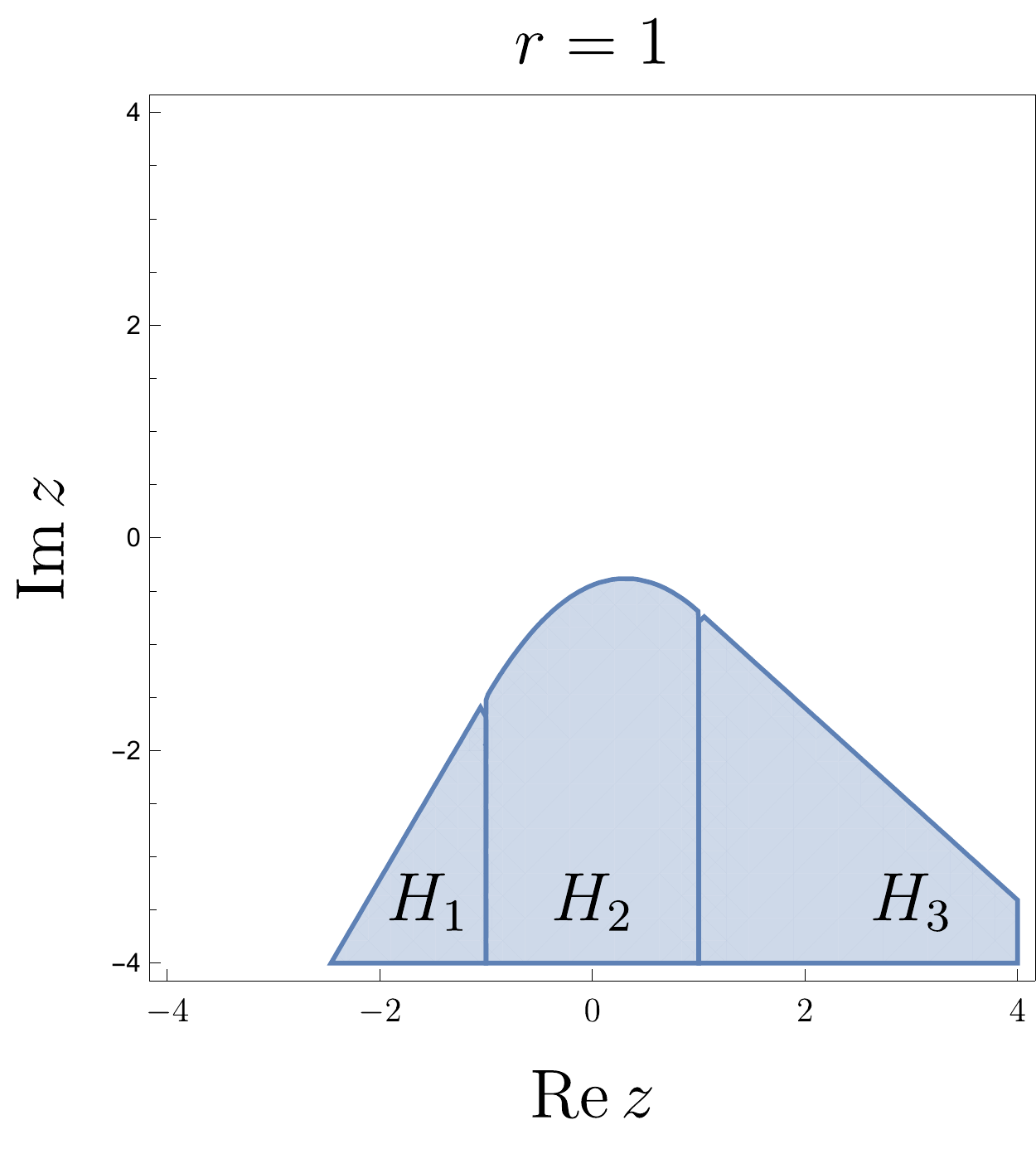}
	\caption{}
\end{subfigure}
	
	\begin{subfigure}[b]{\textwidth}
		\centering
		\includegraphics[scale=.5]{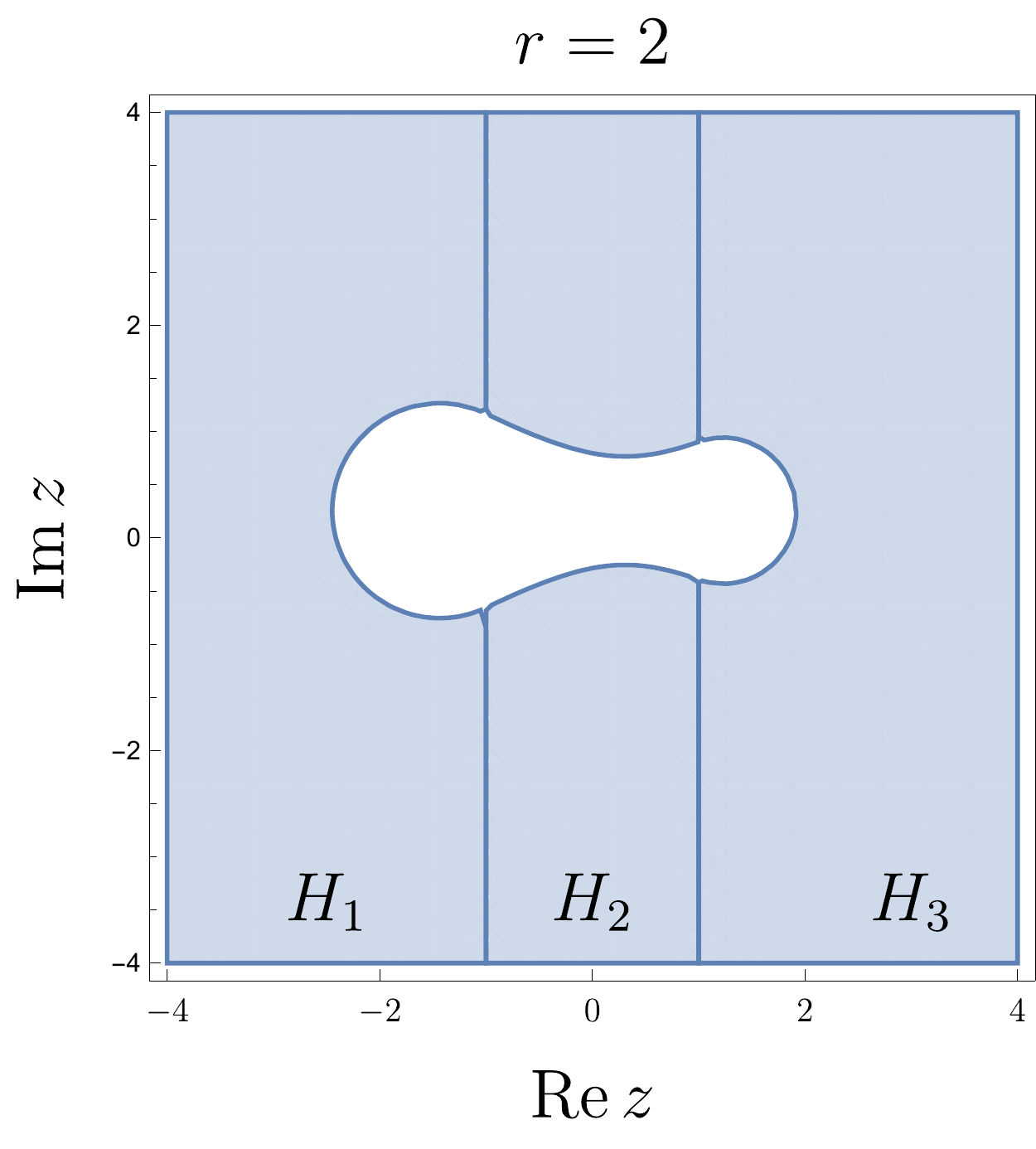}
		\caption{}
	\end{subfigure}
%
	\caption{Representation of the loci $z_k$ for a system for which $z_0=0.308824 -0.764706\,i$, and $R=2.061$.}
	\labfig{fig:loci}
\end{figure}

A conservative choice would be $r=1$ (see \protect{\fig{fig:loci}}b), in which situation $H_1$ and $H_3$ are bounded by straight lines, while $H_2$ is a parabola. To fix ideas, let us assume 
$z_0 = x_0 + i y_0$ with $x_0 \geq 0$ and $y_0 \geq 0$. All other cases being symmetrical. Then the euclidean region $H(1)$ where  $z_1, z_2, \ldots, z_m$ are allowed consists of points $z=x+iy$ subject to the constraints:
\beq x \geq 1 \ \ {\rm and} \ \ {\rm dist}(z,z_0) \leq \ {\rm dist}(z,1), \eeq{a.5h}
union with 
\beq x \in [-1,1] \ \ {\rm and} \ \ (x-x_0)^2 + y_0^2 \leq 2 y_0 y. \eeq{a.5i}
If $y_0 = 0$ then necessarily $x_0 >1$ and $H$ is simply the right-half plane $x > \frac{1+x_0}{2}$, while in the case $y_0>0$, $H(1)$ is the interior of a parabola with vertex at $(x_0, \frac{y_0}{2})$, within the band $|x|\leq 1$, union with the
polygonal region defined by the first distance inequality (in $x \geq 1$).

Now with an input signal of the form \eq{0.4}, with $\Gb_k=\Ga_k/c(\Go_k)$, generating 
the output function $v(t)$ given by \eq{0.5}, \eq{a.1} implies the bound
\beq |v(t_0)-v_0(t_0)|\leq a_0\Ge_m,
\eeq{a.6}
where
\beq v_0(t_0)=a_0 F_\mu(z(\omega_0)) \eeq{a.7}
is the response at time $t_0$ to the single frequency input signal
\beq u_0(t)=e^{-i\Go_0(t-t_0)}/c(\Go_0).  \eeq{a.8}
Of course, because this response $v_0(t)$ is for a single frequency, $v_0(t_0)$ determines $v_0(t)$ for all $t$.

In \protect{\fig{fig:comparison_v_v0}} we depict the response $v_0(t)$ of a given system subject to an input signal at the frequency $\omega_0$ and we compare the value it takes at $t_0=0$ with the value taken by the bounds on the response $v(t)$ of a system having the same values of the moments of the measure but subject to a multiple-frequency signal with amplitudes $\alpha_k$ chosen such that the bounds are extremely tight at $t_0=0$: $v_0(t_0)$ lies, as expected, between the bounds on $v(t)$ at $t=t_0$, 
\begin{figure}[h!]
	\centering
	\begin{subfigure}[pt]{0.5\textwidth}
		\centering
		\includegraphics[width=\textwidth]{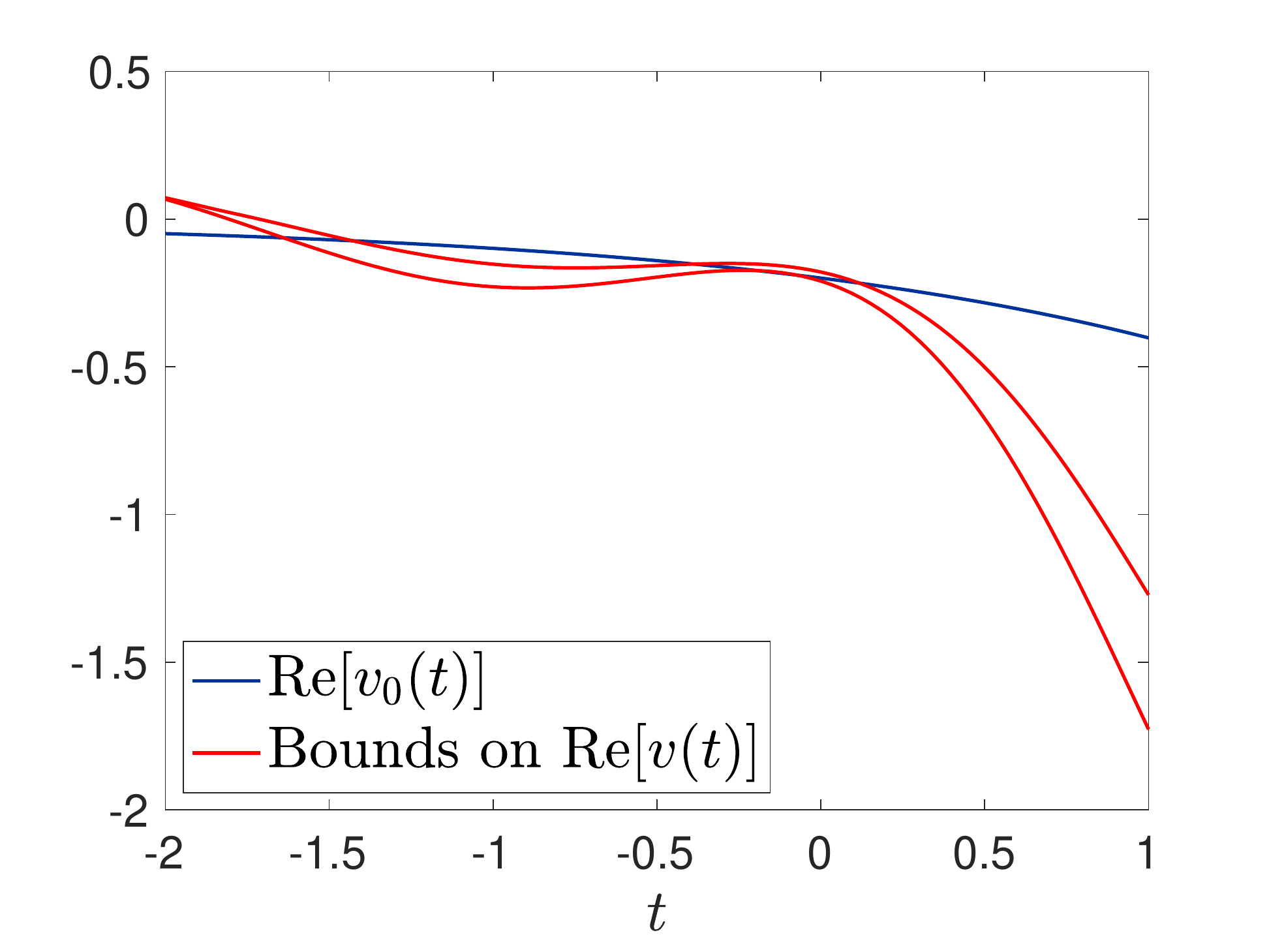}
		\caption{$\displaystyle z=2-\frac{i}{\omega}$}
	\end{subfigure}%
	\begin{subfigure}[pt]{0.5\textwidth}
		\centering
		\includegraphics[width=\textwidth]{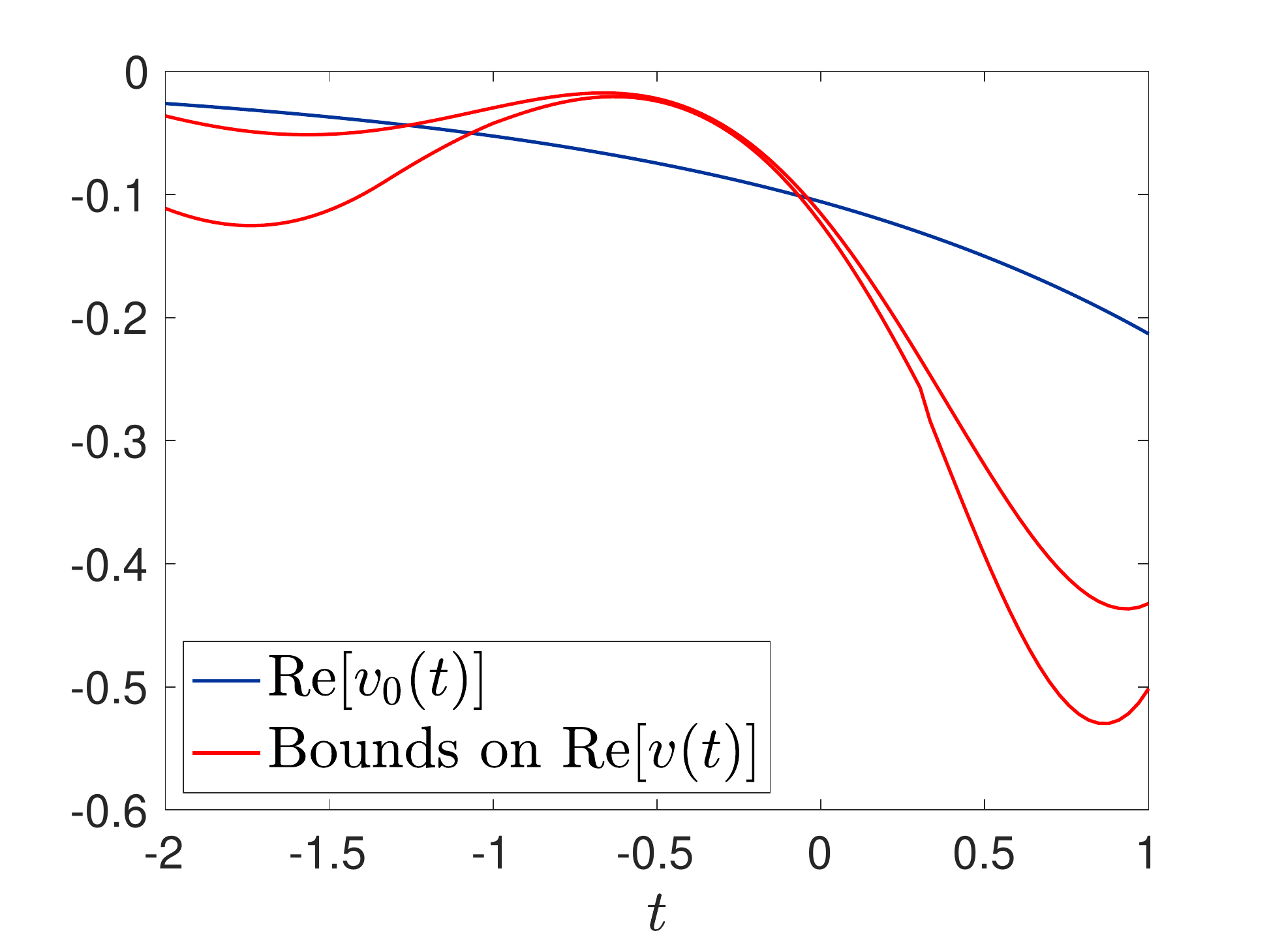}
		\caption{$\displaystyle z=2-\frac{2}{\omega^2}$}
	\end{subfigure}
	\caption{Comparison between the response $v_0(t)$ of a given system with point masses at -0.5 and 0.5, due to an input at the frequency $\omega_0=0.7{i}$, and the upper and lower bounds on the response $v(t)$ of a system having the same value of the moments of the measure $M_i$ ($M_0=1$ and $M_1=0.4$) and subject to an input signal of the type \eqref{0.4}, with $\omega_k$ given by $[1+1i;0.5+0.3i;2+0.5i]$ and coefficients $\beta_k$ chosen accordingly to \eqref{0.5a} and \eqref{a.3a}. Notice that in both cases the value of $v_0(t)$ at $t_0=0$ lies between the bounds on $v(t)$ at $t_0=0$.}
		\labfig{fig:comparison_v_v0}
\end{figure}
\newline

\medskip
\noindent {\bf Remark}
\newline

\smallskip
\noindent
The analysis is easily extended to the case where the response $v_0(t)$ is known for a given $\Go_0$
but one wants to predict the derivative
\beq \frac{v_0(t_0)}{d\Go_0}=a_0\frac{dF_\mu(z)}{dz}\bigg|_{z=z(\omega_0)}
\frac{dz(\omega_0)}{d\Go_0}.
\eeq{a.9}
As
\beq \frac{dF_\mu(z)}{dz}=\int_{-1}^1\frac{d\mu(\Gl)}{(\Gl-z)^2}, \eeq{a.10}
the problem becomes: find complex constants $\Ga_0, \Ga_1,\Ga_2,\ldots, \Ga_n$ such that 
\beq \sup_{\Gl \in [-1,1]} \left| \sum_{j=1}^m \frac{\alpha_j}{\Gl-z_j} +\frac{\alpha_0}{\Gl-z_0} -\frac{1}{(\Gl-z_0)^2}\right|\leq \Ge_m.
\eeq{a.11}
Defining the polynomials $p(\Gl)$ and $q(\Gl)$ as in \eq{1.9} and \eq{1.10} one needs to find $p(\Gl)$ of degree $m-1$ such that
\beq \sup_{\Gl \in [-1,1]} \left|\frac{(\Gl-z_0)^2p(\Gl)-q(\Gl)[1-\Ga_0(\Gl-z_0)]}{(\Gl-z_0)^2q(\Gl)}\right|\leq \Ge_m. 
\eeq{a.12}
We now choose
\beq (\Gl-z_0)^2p(\Gl)=q(\Gl)[1-\Ga_0(\Gl-z_0)]-b_mT_{m-1}(\Gl), \eeq{a.13}
with
\beq b_m=q(z_0)/T_{m-1}(z_0),\quad \Ga_0=\frac{q'(\Gl)-b_mT_{m-1}'(\Gl)}{q(z_0)} \eeq{a.14}
  selected so that the polynomial on the right hand side of \eq{a.13} has a factor of $(\Gl-z_0)^2$, in which $q'(\Gl)=dq(\Gl)/d\Gl$ and
  $T_{m-1}'(\Gl)=dT_{m-1}(\Gl)/d\Gl$. So the residues $\Ga_k$, for $k\ne 0$, are now given by
\beq \Ga_k=-b_m\frac{T_m(z_k)}{(z_k-z_0)^2\prod_{j\ne k}(z_k-z_j)}=-\frac{T_m(z_k)\prod_{j\ne 0}(z_0-z_j)}{T_m(z_0)(z_k-z_0)^2\prod_{j\ne 0,k}(z_k-z_j)},
\eeq{a.15}
where $b_m$ is still given by \eq{a.3} and
\beq \sup_{\Gl \in [-1,1]} |(\Gl-z_0)^2p(\Gl)-q(\Gl)[1-\Ga_0(\Gl-z_0)]|=\sup_{\Gl \in [-1,1]}|b_mT_{m-1}(\Gl)|=|b_m| \eeq{a.16}
so that \eq{a.11} holds with
\beq \Ge_m=\frac{|b_m|}{d_0^2 \inf_{\lambda \in [-1,1]}| q(\lambda)|}. \eeq{a.17}
Apart from an extra factor of $d_0$ this is exactly the same as the formula \eq{a.5}, and so the convergence $\Ge_m\to 0$ as $m\to\infty$ is assured
provided for a positive constant $r<R$, with each $z_j\in H(r)=H_1(r)\cup H_2(r)\cup H_3(r)$.

\section{Framework for a wide variety of time-dependent problems}
\setcounter{equation}{0}
Suppose that, in some Hilbert (or vector space) $\CH$, one is interested in solving for $\BJ$ the equations
\beq \BJ=\BL\BE,\quad \BQ\BJ=\BJ, \quad \BQ\BE=\BE_0, \eeq{5.1}
for a prescribed field $\BE_0$, where $\BL:\CH\to\CH$ is an operator satisfying appropriate boundedness and coercivity conditions, and $\BQ$ is a
selfadjoint projection onto a subspace
$\CS$ of $\CH$, so that both $\BE_0$ and $\BJ$ lie in $\CS$. {Note that we
can rewrite \eq{5.1} as
\beq \BJ=\BL\BE'-\Bs,\quad \BGG_1\BE'=\BE',\quad \BGG_1\BJ=0,
\eeq{5.2aa}
with $\BE'=\BE-\BE_0$, $\Bs=-\BL\BE_0$ being the source term,
and $\BGG_1=\BI-\BQ$ being the
projection onto the orthogonal complement of $\CS$ in $\CH$. These equations
arise in the extended abstract theory of composites and apply to an
enormous plethora of linear continuum equations in physics:
see, for example, the books \cite{Milton:2002:TOC, Milton:2016:ETC}
and the articles \cite{Milton:2020:UPLI, Milton:2020:UPLII, Milton:2020:UPLIII,
  Milton:2020:UPLIV}.} 

{The simplest example is for electrical conductivity (and
equivalent equations), where one has
\beq \Bj'(\Bx)=\BGs(\Bx)\Be(\Bx)-\Bs(\Bx),\quad \BGG_1\Be=\Be,\quad\BGG_1\Bj'=0, \quad\text{with}\quad \BGG_1=\Grad(\Grad^2)^{-1}\Grad\cdot,
\eeq{pt1}
where $\BGs(\Bx)$ is the conductivity tensor,
while $\Div\Bs$, $\Bj=\Bj'+\Bs$, and $\Be$ are the current source, current,
and electric field, and $(\Grad^2)^{-1}$ is the inverse Laplacian
(there is obviously considerable flexibility in the choice of $\Bs(\Bx)$, the
only constraints being square integrability and that
$\Div\Bs$ equals the current source). As current is conserved,
$\Div\Bj=\Div\Bs$, implying $\Div\Bj'=0$, which is clearly equivalent to
$\BGG_1\Bj'=0$. In Fourier space $\BGG_1(\Bk)=\Bk\otimes\Bk/k^2$,
and $\BGG_1\Be=\Be$ implies the Fourier components $\widehat{\Be}(\Bk)$
of $\Be$ satisfy $\widehat{\Be}=-i\Bk(i\Bk\cdot\widehat\Be)/k^2$. So
$\Be$ is the gradient of a potential with  Fourier components
$-i\Bk\cdot\widehat\Be/k^2$.  In antiplane elasticity one takes a 
material with a cross-section in the $(x_1, x_2)$-plane that is
independent of $x_3$, applies
shearing in the $x_3$ direction and observes warping of the cross section.
The displacement $u_3(\Bx)$ in the $x_3$-direction that is associated with this
warping satisfies a conductivity type equation $\Div G\Grad u_3 =\Grad s$,
where $\Grad s$ is a shearing source term (dependent on $(x_1,x_2)$),
$G(x_1,x_2)$ is the shear modulus, and correspondingly
$\Be=-\Grad u_3$ and $\Bj=G\Grad u_3$. The antiplane response
also governs the warping of rods under torsion
for rods that have a non-circular cylindrical shape and are composed of
long fibers aligned with the cylinder axis and embedded in a matrix 
such that the fiber separation is much less than the cylinder circumference.}

One approach to solving \eq{5.1}
is to apply $\BQ$ to both sides of the relation $\BE=\BL^{-1}\BJ$ to obtain $\BE_0=\BQ\BL^{-1}\BQ\BJ$, giving
\beq \BJ=[\BQ\BL^{-1}\BQ]^{-1}\BE_0,
\eeq{5.2}
where the inverse is on the subspace $\CS$. In general the operator $\BL$ depends on the frequency $\Go$ and $\BE_0$ could depend on $\Go$ too. Then the response at this frequency is
\beq \widehat{\BJ}(\omega)=[\BQ(\BL(\Go))^{-1}\BQ]^{-1}\widehat{\BE_0}(\Go). \eeq{5.2a}
We are interested in the response in the time domain when $\widehat{\BE_0}(\Go)=\Gb(\Go)\underline{\BE}_0$ for some complex amplitude $\Gb(\Go)$ and $\underline{\BE}_0\in\CS$
does not depend on $\Go$. In particular, for a sum of a finite
number of (possibly complex) frequencies in the time domain the input
signal is
\beq \BE_0(t)=\sum_{k=1}^me^{-i\Go_k(t-t_0)}\widehat{\BE_0}(\Go_k)=\sum_{k=1}^m\Gb_ke^{-i\Go_k(t-t_0)}\underline{\BE}_0\quad\text{with}\quad\Gb_k=\Gb(\Go_k).
\eeq{5.2b}
The resulting field $\BJ(t)$ is then
\beq \BJ(t)=\sum_{k=1}^m\Gb_k e^{-i\Go_k(t-t_0)}[\BQ(\BL(\Go))^{-1}\BQ]^{-1}\underline{\BE}_0,
  \eeq{5.2c}
and we want this to have a simple approximate formula at time $t_0$. 

To make progress we use another approach to solving \eq{5.1}. {We introduce a ``reference medium''
  $\BL_0=c_0\BI$ where the real constant $c_0$ is chosen so that $\BL-\BL_0$ is coercive}
  and introduce the so-called ``polarization field''
\beq \BG=(\BL-\BL_0)\BE=(\BL-c_0\BI)\BE =\BJ-c_0\BE. \eeq{5.3}
Applying the projection $\BI-\BQ$ to this equation gives
\beq (\BI-\BQ)\BG=-c_0(\BE-\BE_0)=c_0\BE_0-c_0(\BL-c_0\BI)^{-1}\BG, \eeq{5.4}
and solving this for $\BG$ yields
\beq \BG=c_0[(\BI-\BQ)+c_0(\BL-c_0\BI)^{-1}]^{-1}\BE_0. \eeq{5.5}
Finally, applying $\BQ$ to both sides gives
\beq \BJ=c_0\left\{\BQ+\BQ[(\BI-\BQ)+c_0(\BL-c_0\BI)^{-1}]^{-1}\BQ\right\}\BE_0. \eeq{5.6}
By comparing \eq{5.2} and \eq{5.5} we have
\beqa [\BQ\BL^{-1}\BQ]^{-1}& = &c_0\BQ +c_0\BQ[(\BI-\BQ)+c_0(\BL-c_0\BI)^{-1}]^{-1}\BQ\nonum
& = &  c_0\left\{\BQ-2\BQ[\BGY-(\BL+c_0\BI)(\BL-c_0\BI)^{-1}]^{-1}\BQ\right\},
\eeqa{5.7}
where $\BGY=2\BQ-\BI$ has eigenvalues $\pm 1$. It is not obvious at all that the right hand side of \eq{5.7}
is independent of $c_0$ but the preceding derivation shows this. { This type of solution using a reference
medium $\BL_0$ (that need not be proportional to $\BI$) is well known in the theory of composites:
see, for example Chapter 14 of \cite{Milton:2002:TOC},
\cite{Willis:1981:VRM}, and references therein.}

Now assume $\BL$ takes the form
\beq \BL=c_1\BP+c_2(\BI-\BP), \eeq{5.8}
where $\BP$ is a projection operator onto a subspace $\CP$ of $\CH$. {In
the theory of composites for two phase composites one frequently has
\beq \BL=c_1\BI\Gc(\Bx)+c_2\BI(1-\Gc(\Bx)), \eeq{5.8a}
where the characteristic function $\Gc(\Bx)$ is $1$
in phase 1 and 0 in phase 2, and $c_1$ and $c_2$ could be
the material moduli. For the antiplane elasticity problem one has $c_1=G_1$
and $c_2=G_2$, where $G_1$ and $G_2$ are the shear moduli of the phases.} 
We take the limit $c_0\to c_2$ and then \eq{5.7} becomes
\beq [\BQ\BL^{-1}\BQ]^{-1}=c_2\BQ+2c_2\BQ\BP[\BP\BGY\BP-z\BP]^{-1}\BP\BQ,
\eeq{5.9}
where the operator inverse is to be taken on the subspace $\CP$ and
\beq z=\frac{c_1+c_2}{c_1-c_2}. \eeq{5.10}
Note that $\BP\BGY\BP$, like $\BGY$, has norm at most $1$.
In general, the two moduli $c_1$ and $c_2$ depend on the frequency $\Go$ and hence $z$ defined by \eq{5.10} will also, i.e. $z=z(\Go)$. Given an input field of the form \eq{5.2b}
and letting
\beq \BJ_2(t)=\BQ\sum_{k=1}^m \Gb_kc_2(\Go_k)e^{-i\Go_k(t-t_0)}\underline{\BE}_0 \eeq{5.12}
denote the response when $\BP=0$, i.e. when $\BL(\Go)=c_2(\Go)\BI$, the corresponding output field can be taken to be
\beq \Bv(t)=\BJ(t)-\BJ_2(t)=\BQ\sum_{k=1}^m\Ga_k e^{-i\Go_k(t-t_0)} 2\BP[\BP\BGY\BP-z_k\BP]^{-1}\BP\underline{\BE}_0,
\eeq{5.12a}
with
\beq z_k=z(\Go_k)=\frac{c_1(\Go_k)+c_2(\Go_k)}{c_1(\Go_k)-c_2(\Go_k)}, \quad \quad \Ga_k=\Gb_k c_2(\Go_k),
\eeq{5.13}
and we arrive back at the problem we have been studying. In particular, with constants $\Ga_k$ given by \eq{1.c} the inequality $\eq{0.10}$ with $n=0$ implies
\beq |\BJ(t_0)-\BJ_2(t_0)-2\BQ\BP\underline{\BE}_0|\leq 4|\BP\underline{\BE}_0|/(2d_{\rm min})^m.
\eeq{5.14}

Alternatively, we could have chosen $c_0=c_1$ and let
\beq \BJ_1(t)=\BQ\sum_{k=1}^m \Gb_kc_1(\Go_k)e^{-i\Go_k(t-t_0)}\underline{\BE}_0 \eeq{5.15}
denote the response when $\BP=\BI$, i.e. when $\BL(\Go)=c_1(\Go)\BI$. Then, similarly to \eq{5.12a}, we would have
\beq \BJ(t)-\BJ_1(t)=\BQ\sum_{k=1}^m\Ga_k e^{-i\Go_k(t-t_0)} 2\BP_\perp[(\BP_\perp\BGY\BP_\perp+z_k\BP_\perp]^{-1}\BP_\perp\underline{\BE}_0,
\eeq{5.16}
where $z_k$ is still given by \eq{5.13},
but now with $\Ga_k=\Gb_k c_1(\Go_k)$,  where $\BP_\perp=\BI-\BP$ is the projection onto the subspace perpendicular to $\CP$.
The problem, with $n=0$ and with the same choice of coefficients $\Ga_k$, requires a different
input signal, i.e. a different choice of the $\Gb_k$ given by $\Gb_k=\Gb_k/c_1(\Go_k)$, to ensure that
\beq |\BJ(t_0)-\BJ_1(t_0)+2\BQ\BP_\perp\underline{\BE}_0|\leq 4|\BP\underline{\BE}_0|/(2d_{\rm min})^m.
\eeq{5.17}

\section{Framework in the context of the theory of composites and its generalizations}
\setcounter{equation}{0}
In the theory of composites and its generalizations, one can identify a subspace of $\CS$ that we call $\CU$ of ``source free'' fields, and
we may wish to confine $\BE_0$ to this subspace. Then \eq{5.1} can be rewritten as
\beq \BJ=\BL\BE,\quad \BGG_2\BE=0, \quad \BGG_1\BJ=0,\quad \BGG_0\BE=\BE_0, \eeq{5.18}
where $\BGG_0$ is the projection onto $\CU$, $\BGG_1$ is the projection onto $\CE$, defined as the orthogonal complement of $\CS$, and $\BGG_2$ is the projection onto $\CJ$,
defined as the orthogonal complement of $\CU$ in the subspace $\CS$. Then $\BQ=\BGG_0+\BGG_2$ and the Hilbert space $\CH$ has the decomposition
\beq \CH=\CU\oplus\CE\oplus\CJ, \eeq{5.19}
and the projections onto these three subspaces are respectively $\BGG_0$, $\BGG_1$, and $\BGG_2$. 

In particular, as observed independently in Sections 2.4 and 2.5 of \cite{Grabovsky:2016:CMM}, and
in Chapter 3 of \cite{Milton:2016:ETC}, the Dirichlet-Neumann problem can be reformulated
as a problem in the theory of composites. In the simplest case of electrical conductivity,
where one has an inclusion $D$
(not necessarily simply connected) of (isotropic) conductivity $c_1$ in a simply connected body $\GO$ having smooth boundary, with $c_2$ being the (isotropic)
conductivity of $\GO\setminus D$, we may take
$\CH$ as the space of vector fields that are square integrable with the usual normalized $L^2$ inner product,
\beq (\BA_1,\BA_2)=\frac{1}{|\GO|}\int_\GO \BA_1(\Bx)\cdot\overline{\BA_2(\Bx)}\,d\Bx, \eeq{5.19a}
where $|\GO|$ is the volume of $\GO$, 
and take
\begin{itemize}
\item  $\CU$ to consist of gradients of harmonic fields $\Bu_0=-\Grad V$ with $\Grad^2V=0$ in $\GO$,
\item  $\CE$ to consist of gradients $\Be=-\Grad V$ with $V=0$ on the boundary $\Md\GO$ of $\GO$,
\item  $\CJ$ to consist of divergent free vector fields $\Bj$ with $\Div\Bj=0$ and $\Bj\cdot\Bn=0$ on $\Md\GO$, where $\Bn$ is the outwards normal to $\Md\GO$.
\end{itemize}
The conductivity of the body may be identified with $\BL$ given by \eq{5.8} where $\BP$ is the projection onto those fields that are zero outside $D$. 
As we are considering time-dependent problems in the quasistatic limit, where the body is small compared to the wavelength and attenuation lengths of
electromagnetic waves at the frequencies $\Go_k$, the moduli $c_1$ and $c_2$ and the fields are typically complex and frequency-dependent. The fields in $\CU$ can be identified either by the values that $V$
takes on the boundary $\Md\GO$ or by the values that the flux $\Bn\cdot\Grad V$
takes on the boundary $\Md\GO$. Thus the equations
\eq{5.18} are nothing other than the Dirichlet problem in the body $\GO$,
\beq  \Bj=\BL\Be,\quad \Be=-\Grad V, \quad \Div\Bj=0,\quad \Be_0=-\Grad V_0,\quad \Grad^2V_0=0,\quad V=V_0,\quad \text{on}\quad\Md\GO, \eeq{5.20}
and the mapping from $\BGG_0\Be$ to $\BGG_0\Bj$ is nothing other than the Dirichlet to Neumann map giving $\Bn\cdot\Bj$ in terms of $V$ on $\partial\Omega$.

For periodic two-phase conducting composites, with unit cell $\GO$, the framework is similar. We take
$\CH$ as the space of vector fields that are $\GO$-periodic with the usual normalized $L^2$ inner product, given by \eq{5.19a}, and take
\begin{itemize}
\item  $\CU$ to consist of gradients of constant fields $\Bu_0$ (that do not depend on $\Bx$),
\item  $\CE$ to consist of gradients $\Be=-\Grad V$ with $V$ being an $\GO$-periodic potential,
\item  $\CJ$ to consist of $\GO$-periodic divergent free vector fields $\Bj$ with $\Div\Bj=0$, having zero average over $\GO$.
\end{itemize}
The conductivity of the body may be identified with $\BL$ given by \eq{5.8} where $\BP$ is the projection onto those fields in $\CH$ that are zero outside
phase 1, and $c_1$ is the (isotropic) conductivity of phase 1 while $c_2$ is the (isotropic) conductivity of phase 2.
\newline

\medskip
\noindent {\bf Remark}
\newline

\smallskip
\noindent
More generally, the conductivity in the periodic composite
could be anisotropic, with the conductivity tensor having the special form
\beq  \BL(\Go)=c_1(\Go)\BL_0\BP+c_2(\Go)\BL_0(\BI-\BP), \eeq{5.20a}
where $\BL_0$ is a constant positive definite tensor. As $\BL_0$ commutes with $\BGG_0$ and $\BP$,
we can define new orthogonal spaces
\beq \CE'=\BL_0^{1/2}\CE,\quad\CJ'=\BL_0^{-1/2}\CJ,\quad \CU'=\BL_0^{1/2}\CU=\BL_0^{-1/2}\CU=\CU, \eeq{5.20b}
and rewrite \eq{5.18} in the form
\beq \BJ'=\BL'\BE',\quad \BGG_2'\BE'=0, \quad \BGG_1'\BJ'=0,\quad \BGG_0'\BE'=\BE_0', \eeq{5.20c}
where
\beqa \BJ'& = & \BL_0^{-1/2}\BJ,\quad \BE'=\BL_0^{1/2}\BE,\quad \BE'_0=\BL_0^{1/2}\BE_0, \nonum
\BL'& = & \BL_0^{-1/2}\BL\BL_0^{-1/2}=c_1(\Go)\BP+c_2(\Go)(\BI-\BP),
\eeqa{5.20d}
and
\beq \BGG_0'=\BGG_0,\quad\BGG_1'=\BL_0^{-1/2}\BGG_1(\BGG_1\BL_0\BGG_1)^{-1},\quad \BGG_2'=\BI-\BGG_1'-\BGG_2'
\eeq{5.20e}
are the projections onto $\CU'=\CU$, $\CE'$, and $\CJ'$, in which the inverse in the formula for $\BGG_1'$ is to be taken on the subspace $\CE$. As $\BL'$ now takes
the same form as \eq{5.8} we are back to the same problem.

Similarly, in a body where the  conductivity tensor has the special form \eq{5.20a} we may take
\begin{itemize}
\item  $\CU'$ to consist of gradients of fields $\Bu_0=-\BL_0^{1/2}\Grad V$ with $\Div\BL_0\Grad V=0$ in $\GO$,
\item  $\CE'$ to consist of fields $\Be'=-\BL_0^{1/2}\Grad V$ with $V=0$ on the boundary $\Md\GO$ of $\GO$,
\item  $\CJ'$ to consist of fields $\Bj'$ with $\Div\BL_0^{1/2}\Bj'=0$ and $(\BL_0^{1/2}\Bj')\cdot\Bn=0$ on $\Md\GO$, where $\Bn$ is the outwards normal to $\Md\GO$
\end{itemize}
as our three orthogonal subspaces.   Letting $\BGG_0$, $\BGG_1$, and $\BGG_2$ denote the projections onto these three subspaces, respectively, and 
setting $\BL'=c_1(\Go)\BP+c_2(\Go)(\BI-\BP)$, the equations \eq{5.20b} hold and we may proceed as before. 

\section{Application to solving the Calderon problem with time varying fields}
\setcounter{equation}{0}
\labsect{Cald}

Let us now use ideas from the Calderon problem to solve the inverse problem of finding the inclusion $D$ from boundary measurements on $\partial\Omega$.  
With $\GO$ being a three-dimensional body, we can take
\beq V_0=e^{i\Bk\cdot\Bx}\quad \text{with}\quad k_1, k_2\quad\text{real and}\quad k_3=i\sqrt{k_1^2+k_2^2}, \eeq{5.21}
where the last condition implies $\Bk\cdot\Bk=0$ which ensures that $V_0$ is harmonic. Then \eq{5.14} implies
\beq (\BJ(t_0)-\BJ_1(t_0)+2\BQ\BP\underline{\BE}_0,\Grad e^{i\Bk'\cdot\Bx})\leq 4|\Bk||\Bk'|/(2d_{\rm min})^m \eeq{5.22}
for all real or complex $\Bk'$. We now choose $\Bk'$ with 
\beq k_3'=-k_3,\quad k_1', k_2'\,\,\text{real and with}\quad (k_1')^2+(k_1')^2=k_1^2+k_2^2 \eeq{5.23}
  to ensure that $e^{i\Bk'\cdot\Bx}$ is harmonic and so that
  \beqa (2\BQ\BP\underline{\BE}_0,\Grad e^{i\Bk'\cdot\Bx}) & = & 2(\BP\underline{\BE}_0,\BQ\Grad e^{i\Bk'\cdot\Bx})=2(\BP\underline{\BE}_0,\Grad e^{i\Bk'\cdot\Bx}) \nonum
  & = & 2(k_1k'_1+k_2k_2'-k_1^2-k_2^2)\frac{1}{|\GO|}\int_D e^{i(k_1-k'_1)x_1+i(k_2-k'_2)x_2}\,d\Bx \nonum &~&
  \eeqa{5.23a}
  only depends on the Fourier coefficients of the characteristic function associated with $D$. Then, using integration by parts,
  \beq (\BJ(t_0)-\BJ_1(t_0),\Grad e^{i\Bk'\cdot\Bx})=\frac{1}{|\GO|}\int_{\Md\GO}[\BJ(t_0)-\BJ_1(t_0)]\cdot\Bn\, e^{i\Bk'\cdot\Bx}\,dS,
  \eeq{5.24}
where $\BJ(t_0)\cdot\Bn$ can be measured, while $\BJ_1(t_0)\cdot\Bn$ can be computed. As there is nothing special about the $x_3$ axis
  we may rotate the cartesian coordinates to get estimates of other Fourier coefficients of the characteristic function associated with $D$.
  We may also take $\underline{\BE}_0$ as constant and replace $\Grad e^{i\Bk'\cdot\Bx}$ by $\underline{\BE}_0$ to get
  \beq (\BJ(t_0)-\BJ_1(t_0)+2\BQ\BP\underline{\BE}_0,\underline{\BE}_0)=(\BJ(t_0)-\BJ_1(t_0),\underline{\BE}_0)+|\underline{\BE}_0|^2|D|/|\GO| \leq 4|\underline{\BE}_0|^2/(2d_{\rm min})^m,
  \eeq{5.24a}
  thus giving an estimate of the volume fraction $|D|/|\GO|$ that $D$ occupies in the body (i.e., the Fourier coefficient at $\Bk=0$).

  With $\GO$ being a two-dimensional body, the situation is similar. We take
  \beq k_1\,\, \text{real and}\,\, k_2=ik_1,\quad k_1'=-k_1,\quad k_2'=-ik_1,
  \eeq{5.25}
  and \eq{5.23a} is replaced by
  \beq (2\BQ\BP\underline{\BE}_0,\Grad e^{i\Bk'\cdot\Bx})=-4k_1^2\int_D e^{2ik_1x_1}\,d\Bx,
  \eeq{5.26}
  while \eq{5.24} and \eq{5.24a} still hold. Again we approximately recover the Fourier coefficients of the characteristic function associated with $D$
  from measurements of $\BJ(t_0)\cdot\Bn$ and computations of $\BJ_1(t_0)\cdot\Bn$.

  In the usual Calderon problem one solves the inverse problem
  by taking $|\Bk|$ to be very large, according to the so-called complex geometric optics approach \cite{Sylvester:1987:GUT}. Here we see that there is no need to take  $|\Bk|$ to be very large if we
  allow time dependent applied fields. For electromagnetism in non-magnetic media
  the measurements are difficult as the time response is typically extremely rapid
  (From Table 7.7.1 in \cite{Haus:1989:EFE} we see that electromagnetic relaxation times in seconds for copper, distilled water, corn-oil, and mica are $1.5\times 10^{-19}$, $3.6\times 10^{-6}$,
  $0.55$, and $5.1\times 10^4$ respectively, and measurements would need to be taken on these time scales).
  On the other hand, for the equivalent magnetic permeability, fluid permeability, or
  antiplane elasticity problems the relaxation times are much more reasonable
  \cite{Balanda:2013:ASS, Johnson:1987:TDP, Lakes:1996:VBI}
  and measurements in the time domain become feasible. Even in electrical
  systems one can get long relaxation times, such as the time to charge a
  capacitor.


\medskip
\noindent {\bf Remark}
\newline

\smallskip
\noindent

  Instead of taking $\BE_0(t)=\BGG_0\BE(t)$ and $\BJ(t)$ as our input and output fields, one could take $\BJ_0(t)=\BGG_0\BJ(t)$ and $\BE(t)$ as our input and output fields. Then one has
  \beq \BE=\BL^{-1}\BJ,\quad \BGG_1\BJ=0, \quad \BGG_2\BE=0,\quad \BGG_0\BJ=\BJ_0, \eeq{5.27}
  which is exactly of the same form as \eq{5.18}, but with $\BL$ replaced by $\BL^{-1}$ and the roles of $\BGG_1$, $\BGG_2$, and $\BE$ and $\BJ$, and $\BE_0$ and $\BJ_0$ interchanged.
  So all the preceding analysis immediately applies to this dual problem too. 
  
\section{Generalizations}
\setcounter{equation}{0}
\labsect{Gen}
In many problems of interest the fields in $\CH$ take values in a, say, $s$-dimensional tensor space $\CT$
and the operator $\BL:\CH\to\CH$ in \eq{5.1}, appropriately defined, is frequency dependent
with the properties that
\begin{itemize}
\item $\BL(\Go)$ is an analytic function of $\Go$ in the upper half plane $\Imag(\Go)> 0$,
\item $\Imag[\Go\BL(\Go)]\geq 0$ when $\Imag(\Go)> 0$,
\item $\overline{\BL(\omega)}=\BL(-\overline{\omega})$ when $\Imag(\Go)> 0$,
\end{itemize}
where the overline denotes complex conjugation. 
By appropriately defined we mean that $\BL(\Go)$ (and accordingly $\BJ$) may need to be multiplied
by a function of $\Go$, for example $i$, $\Go$, or $i\Go$, to achieve these properties. 
In the case of materials where $\BL$ acts locally in real space, i.e. if $\BQ=\BL\BP$,
then $\BQ(\Bx)=\BL(\Bx)\BP(\Bx)$ for some $\BL(\Bx)$,
the first property is a consequence of causality, the second a consequence
of passivity (that the material does not generate
energy - see, for example, \cite{Welters:2014:SLL}), and the third a consequence of $\BL(\Go)$ being the Fourier transform of a real kernel.
It follows that $\BL$ is an analytic function of $-\Go^2$ with spectrum on the negative real $-\Go^2$ axis
(corresponding to real values of $\Go$) having the implied properties that
\begin{itemize}
  \item $\Imag(\BL)\geq 0$ when $\Imag(-\Go^2)\leq 0$,
 \item $\BL$ is real and $\BL \geq 0$ when $\Go^2$ is real and $-\Go^2\geq 0$.
\end{itemize}
In other words, $\BL(\Go)$ is an operator-valued Stieltjes function of $-\Go^2$. The operator $\BB=[\BQ\BL^{-1}\BQ]^{-1}$
entering \eq{5.2} has the property that it is an analytic function of $\BL$ with
\beqa \Imag(\BB) & \geq & 0\,\,\text{when}\,\,\Imag(\BL) \geq 0, \nonum
\BB\text{ is real and } \BB & \geq & 0\,\,\text{when}\,\,\BL \text{ is real and }\BL \leq 0.
\eeqa{6.2}
Hence, the Stieltjes properties of $\BL$ as a function of $-\Go^2$ pass to those of $\BB$ as a function of $-\Go^2$:
\beqa \Imag(\BB) & \geq & 0\,\,\text{when}\,\,\Imag(-\Go^2)\leq 0, \nonum
\BB\text{ is real and } \BB & \geq & 0\,\,\text{when}\,\,\Go^2 \text{ is real and } -\Go^2 \geq 0.
\eeqa{6.3}
Introducing
\beq z=\frac{\Go^2-c}{\Go^2+c}=1-\frac{2c}{\Go^2+c}, \eeq{6.4}
for some real $c>0$, ensures that the spectrum of $\BB(z)$ is on the interval $[-1,1]$ and
\beqa \Imag(\BB(z))& \geq &  0\,\,\text{when}\,\,\Imag(z)\geq 0,\nonum
\BB\text{ is real and } \BB & \geq & 0\,\,\text{when}\,\,z \text{ is real and }z>1\text{ or }z<-1.
\eeqa{6.5}
Note that this choice of $z$ is quite different to that in \eq{5.10}, and not restricted to two-phase
composites. 
Thus, $\BB(z)$ has the integral representation
\beq \BB(z)=\BB_0+\int_{-1}^1\frac{d\BM(\Gl)}{\Gl-z},
\eeq{6.6}
where $\BB_0$ is a positive definite operator and $d\BM(\Gl)$ is a positive definite real operator-valued measure,
satisfying the constraint
\beq \int_{-1}^1\frac{d\BM(\Gl)}{1-\Gl}\leq\BB_0.
\eeq{6.7}
To begin, suppose we are only interested in the quadratic form $(\BB\BE_0,\BE_0)$ associated with $\BB$. Then,
\beq  (\BB(z)\BE_0,\BE_0)=k_0\left[1+\int_{-1}^1\frac{(1-\Gl)d\Gn(\Gl)}{\Gl-z}\right]
=k_0\left\{1+\int_{-1}^1\left[-1+\frac{1-z}{\Gl-z}\right]d\Gn(\Gl)\right\},
\eeq{6.8}
where $k_0=(\BB_0\BE_0,\BE_0)$ is real and positive and
\beq d\Gn(\Gl)=(d\BM(\Gl)\BE_0,\BE_0)/[k_0(1-\Gl)] \eeq{6.8a}
is a positive real valued measure, satisfying the constraint
\beq \int_{-1}^1 d\Gn(\Gl)\leq 1.
\eeq{6.9}
Note that $k_0$ can be identified with $(\BB(z)\BE_0,\BE_0)$ in the limit $z\to\infty$, i.e. as $\Go\to i\sqrt{c}$.

If we are interested in finding complex coefficients $\xi_k$, $k=1,2\ldots, m$, such that 
\beqa &~& (\BB(z_0)\BE_0,\BE_0)-\sum_{k=1}^m \xi_k(\BB(z_k)\BE_0,\BE_0) \nonum
&~&\quad =k_0\left\{(1-\sum_{k=1}^m\xi_k)\left[1-\int_{-1}^1d\Gn(\Gl)\right]+\int_{-1}^1\left[\frac{1-z_0}{\Gl-z_0}-\sum_{k=1}^m\frac{\xi_k(1-z_k)}{\Gl-z_k}\right]d\Gn(\Gl)\right\} \nonum
&~&
\eeqa{6.10}
is small, we require that
\beq \sup_{\Gl\in [-1,1]}\left|\frac{1}{\Gl-z_0}-\sum_{k=1}^m\frac{\xi_k(1-z_k)/(1-z_0)}{\Gl-z_k}\right|\leq \Ge_m.
\eeq{6.11}
In particular, with $\Gl=1$, this implies
\beq |1-\sum_{k=1}^m\xi_k|\leq |1-z_0|\Ge_m,
\eeq{6.12}
and so we obtain
\beq(\BB(z_0)\BE_0,\BE_0)-\sum_{k=1}^m \xi_k(\BB(z_k)\BE_0,\BE_0)\leq 2k_0|1-z_0|\Ge_m.
\eeq{6.13}
By setting $\Ga_k=\xi_k(1-z_k)/(1-z_0)$ we see  this is exactly the problem encountered in Section 6, and we may take the coefficients $\Ga_k$ to be given by \eq{a.3a}. The motivation
for studying this problem is that the response at special frequencies can sometimes directly reveal information about the geometry. This is the case for elastodynamics in the quasistatic
limit when only two materials are present. The material parameters are the bulk moduli $\Gk_1(\Go)$, $\Gk_2(\Go)$ and shear moduli $\Gm_1(\Go)$, $\Gm_2(\Go)$ of the two phases. It
may happen that $\Gm_1(\Go_0)=\Gm_2(\Go_0)$ for certain complex frequencies $\Go_0$ and if $\Gk_1(\Go_0)\ne\Gk_2(\Go_0)$ the response at frequency $\Go_0$ can reveal the volume fraction
of phase 1 in a composite, or more generally in a two-phase body.
\newline

\medskip
\noindent {\bf Remark 1}
\newline

\smallskip
\noindent
It is not much more difficult to treat bilinear forms. Then we have
\beqa 4(\BB(z)\BE_0,\BE_0') &= &
(\BB(z)(\BE_0+\BE_0'),\BE_0+\BE_0')-(\BB(z)(\BE_0-\BE_0'),\BE_0-\BE_0')\nonum
  & = &k_0^{(1)}\left\{1+\int_{-1}^1\left[-1+\frac{1-z}{\Gl-z}\right]d\Gn_1(\Gl)\right\} \nonum
  &~& -k_0^{(2)}\left\{1+\int_{-1}^1\left[-1+\frac{1-z}{\Gl-z}\right]d\Gn_2(\Gl)\right\},
\eeqa{6.14}
where
\beq k_0^{(1)}=(\BB_0(\BE_0+\BE_0'),\BE_0+\BE_0'),\quad
k_0^{(2)}=(\BB_0(\BE_0-\BE_0'),\BE_0-\BE_0')
\eeq{6.15}
are both real and positive, while 
\beqa d\Gn_1(\Gl)& = & (d\BM(\Gl)(\BE_0+\BE_0'),\BE_0+\BE_0')/[k_0^{(1)}(1-\Gl)],\nonum
d\Gn_2(\Gl)& = &(d\BM(\Gl)(\BE_0-\BE_0'),\BE_0-\BE_0')/[k_0^{(2)}(1-\Gl)]
\eeqa{6.16}
are positive real valued measures, satisfying the constraints that
\beq \int_{-1}^1 d\Gn_1(\Gl)\leq 1, \quad \int_{-1}^1 d\Gn_2(\Gl)\leq 1.
\eeq{6.17}
We seek complex coefficients $\xi_k$, $k=1,2\ldots, m$, such that 
\beqa &~& (\BB(z_0)\BE_0,\BE_0')-\sum_{k=1}^m \xi_k(\BB(z_k)\BE_0,\BE_0') \nonum
&=&\quad k_0^{(1)}\left\{(1-\sum_{k=1}^m\xi_k)\left[1-\int_{-1}^1d\Gn_1(\Gl)\right]+\int_{-1}^1\left[\frac{1-z_0}{\Gl-z_0}-\sum_{k=1}^m\frac{\xi_k(1-z_k)}{\Gl-z_k}\right]d\Gn_1(\Gl)\right\}/4 \nonum
&~&-k_0^{(2)}\left\{(1-\sum_{k=1}^m\xi_k)\left[1-\int_{-1}^1d\Gn_2(\Gl)\right]+\int_{-1}^1\left[\frac{1-z_0}{\Gl-z_0}-\sum_{k=1}^m\frac{\xi_k(1-z_k)}{\Gl-z_k}\right]d\Gn_2(\Gl)\right\}/4
\nonum &~&
\eeqa{6.18}
is small. Using the bounds \eq{6.11} we obtain
\beq \left|(\BB(z_0)\BE_0,\BE_0')-\sum_{k=1}^m \xi_k(\BB(z_k)\BE_0,\BE_0')\right|\leq (k_0^{(1)}+k_0^{(2)})|1-z_0|\Ge_m/2.
\eeq{6.19}
\newline

\medskip
\noindent {\bf Remark 2}
\newline

\smallskip
\noindent
Noting that
\beq \frac{d}{dz}(\BB(z)\BE_0,\BE_0)=k_0\int_{-1}^1\frac{(1-\Gl)d\Gn(\Gl)}{(\Gl-z)^2}
=k_0\int_{-1}^1\frac{1}{\Gl-z}\left[-1+\frac{1-z}{\Gl-z}\right]d\Gn(\Gl),
\eeq{6.20}
we can easily obtain bounds that correlate this derivative at $z_0$ with the values of $(\BB(z_k)\BE_0,\BE_0)$, $k=0,1,2\ldots,m$. We
seek complex constants $\Gg_k$, $k=0,1,2\ldots, m$, such that 
\beqa &~&\left[\frac{d}{dz_0}(\BB(z_0)\BE_0,\BE_0)\right]-\sum_{k=0}^m \xi_k(\BB(z_k)\BE_0,\BE_0) \nonum
&~&\quad =k_0\left\{\left(\sum_{k=0}^m\xi_k\right)\left[1-\int_{-1}^1d\Gn(\Gl)\right]+\int_{-1}^1\left[-\frac{1}{\Gl-z_0}+\frac{1-z_0}{(\Gl-z_0)^2}
    -\sum_{k=0}^m\frac{\xi_k(1-z_k)}{\Gl-z_k}\right]d\Gn(\Gl)\right\} \nonum &~&
\eeqa{6.21}
is small, and this is ensured if
\beq \sup_{\Gl\in [-1,1]}\left|\sum_{k=0}^m\frac{\xi_k(1-z_k)/(1-z_0)}{\Gl-z_k}+\frac{\xi_0+[1/(1-z_0)]}{\Gl-z_0}-\frac{1}{(\Gl-z_0)^2}\right|\leq \Ge_m,
\eeq{6.22}
and $\Ge_m\to 0$ as $m\to\infty$. Observe that \eq{6.22} with $\Gl=1$ implies 
\beq \left|\sum_{k=0}^m\xi_k\right|\leq |1-z_0|\Ge_m.
\eeq{6.23}
Comparing \eq{6.22} with \eq{a.11} we see that we should choose 
\beq \xi_0=\Ga_0-[1/(1-z_0)],\quad \xi_k=\Ga_k(1-z_0)/(1-z_k), \eeq{6.24}
and then, with $b_m$ and coefficients $\Ga_k$ given by \eq{a.14} and \eq{a.15}, \eq{6.22} holds with $\Ge_m$ given by \eq{a.17}. Then
\beq \left|\left[\frac{d}{dz_0}(\BB(z_0)\BE_0,\BE_0)\right]-\sum_{k=0}^m \xi_k(\BB(z_k)\BE_0,\BE_0)\right|\leq 2|k_0||1-z_0|\Ge_m,
\eeq{6.25}
holds, and similarly one has
\beq \left|\left[\frac{d}{dz_0}(\BB(z_0)\BE_0,\BE_0')\right]-\sum_{k=0}^m \xi_k(\BB(z_k)\BE_0,\BE_0')\right|\leq (k_0^{(1)}+k_0^{(2)})|1-z_0|\Ge_m.
\eeq{6.26}
The convergence of $\Ge_m$ to zero as $m\to\infty$ is again ensured provided for a positive constant $r<R$, where $R$ is defined
by \eq{a.5a}, each $z_k\in H(r)=H_1(r)\cup H_2(r)\cup H_3(r)$, where the regions $H_i$, $i=1,2,3$, are given by \eq{a.5g}. The motivation for studying
this problem is that the response may be trivial at certain frequencies $\Go_0$ while the derivative of the response with respect to $\Go$ at $\Go=\Go_0$ directly
reveals some information about the body. This is the case for electromagnetism when only two non-magnetic materials are present (with magnetic permeabilities
$\Gm_1=\Gm_2=\Gm_0$ where $\Gm_0$ is the permeability of the vacuum). It may happen that the electric permittivities of the two phases satisfy $\Gve_1(\Go_0)=\Gve_2(\Go_0)$
for certain complex frequencies $\Go_0$. At this frequency $\Go_0$ the body is homogeneous and its response can be easily calculated. Using perturbation theory
and assuming $d\Gve_1(\Go_0)/d\Go_0\ne d\Gve_2(\Go_0)/d\Go_0$ the derivative of the response with respect to $\Go$ at $\Go=\Go_0$ 
reveals information about the distribution of the two phases in the body.

\section*{Acknowledgements}
GWM and OM are grateful to the National Science Foundation for support through the Research Grants DMS-1211359 and DMS-1814854, and DMS-2008105, respectively. MP was partially supported by a Simons Foundation collaboration grant for mathematicians.




\begin{thebibliography}{10}

\bibitem{Alessandrini:1998:ICP}
Giovanni Alessandrini and Edi Rosset.
\newblock The inverse conductivity problem with one measurement: Bounds on the
  size of the unknown object.
\newblock {\em SIAM Journal on Applied Mathematics}, 58(4):1060--1071, August
  1998.

\bibitem{Allaire:2002:SOH}
Gr{\'e}goire Allaire.
\newblock {\em Shape Optimization by the Homogenization Method}, volume 146 of
  {\em Applied Mathematical Sciences}.
\newblock Springer-Verlag, Berlin, Germany~/ Heidelberg, Germany~/ London, UK~/
  etc., 2002.

\bibitem{Ammari:2007:PMT}
H.~Ammari and H.~Kang.
\newblock {\em Polarization and moment tensors: with applications to inverse
  problems and effective medium theory}, volume 162.
\newblock Springer Science \& Business Media, New York, 2007.

\bibitem{Ammari:2004:RSI}
Habib Ammari and Hyeonbae Kang.
\newblock {\em Reconstruction of Small Inhomogeneities from Boundary
  Measurements}, volume 1846 of {\em Lecture Notes in Mathematics}.
\newblock Springer-Verlag, Berlin, Germany~/ Heidelberg, Germany~/ London, UK~/
  etc., 2004.

\bibitem{Balanda:2013:ASS}
M.~Ba\l{l}anda.
\newblock {AC} susceptibility studies of phase transitions and magnetic
  relaxation: {Conventional}, molecular and low-dimensional magnets.
\newblock {\em Acta Physica Polonica A}, 124(6):964--976, December 2013.

\bibitem{Baratchart:2001:BMA}
L.~Baratchart, V.~A. Prokhorov, and E.~B. Saff.
\newblock Best meromorphic approximation of {M}arkov functions on the unit
  circle.
\newblock {\em Found. Comput. Math.}, 1(4):385--416, 2001.

\bibitem{Bergman:1978:DCC}
David~J. Bergman.
\newblock The dielectric constant of a composite material --- {A} problem in
  classical physics.
\newblock {\em Physics Reports}, 43(9):377--407, July 1978.

\bibitem{Bergman:1982:RBC}
David~J. Bergman.
\newblock Rigorous bounds for the complex dielectric constant of a
  two-component composite.
\newblock {\em Annals of Physics}, 138(1):78--114, 1982.

\bibitem{Bruhl:2003:DIT}
Martin Br{\"u}hl, Martin Hanke, and Michael~S. Vogelius.
\newblock A direct impedance tomography algorithm for locating small
  inhomogeneities.
\newblock {\em Numerische Mathematik}, 93(4):635--654, February 2003.

\bibitem{Calderon:1980:IBV}
Alberto-P. Calder{\'o}n.
\newblock On an inverse boundary value problem.
\newblock In {\em {Seminar on Numerical Analysis and its Applications to
  Continuum Physics: 24 a 28 de Mar{\c{c}}o 1980}}, volume~12 of {\em
  Cole{\c{c}}{\~a}o Atas}, pages 65--73. Sociedade Brasiliera de
  Mathem{\'a}tica, Rio de Janeiro, Brazil, 1980.

\bibitem{Capdeboscq:2003:OAE}
Yves Capdeboscq and Michael~S. Vogelius.
\newblock Optimal asymptotic estimates for the volume of internal
  inhomogeneities in terms of multiple boundary measurements.
\newblock {\em Mathematical Modelling and Numerical Analysis = Modelisation
  math{\'e}matique et analyse num{\'e}rique: $M^2AN$}, 37(2):227--240,
  March\slash April 2003.

\bibitem{Carini:2015:VFL}
A.~Carini and Ornella Mattei.
\newblock Variational formulations for the linear viscoelastic problem in the
  time domain.
\newblock {\em European Journal of Mechanics, A, Solids}, 54:146--159,
  November\slash December 2015.

\bibitem{Cherkaev:2000:VMS}
Andrej~V. Cherkaev.
\newblock {\em Variational Methods for Structural Optimization}, volume 140 of
  {\em Applied Mathematical Sciences}.
\newblock Springer-Verlag, Berlin, Germany~/ Heidelberg, Germany~/ London, UK~/
  etc., 2000.

\bibitem{Gibiansky:1993:EVM}
Leonid~V. Gibiansky and Graeme~W. Milton.
\newblock On the effective viscoelastic moduli of two-phase media. {I}.
  {Rigorous} bounds on the complex bulk modulus.
\newblock {\em Proceedings of the Royal Society of London. Series A,
  Mathematical and Physical Sciences}, 440(1908):163--188, January 1993.

\bibitem{Golden:1983:BEP}
Kenneth~M. Golden and George~C. Papanicolaou.
\newblock Bounds for effective parameters of heterogeneous media by analytic
  continuation.
\newblock {\em Communications in Mathematical Physics}, 90(4):473--491, 1983.

\bibitem{Gonchar:1978:MTM}
A.~A. Gon\v{c}ar and Giermo Lopes~L.
\newblock Markov's theorem for multipoint {P}ad\'{e} approximants.
\newblock {\em Mat. Sb. (N.S.)}, 105(147)(4):512--524, 639, 1978.

\bibitem{Grabovsky:2016:CMM}
Yury Grabovsky.
\newblock {\em Composite Materials: Mathematical Theory and Exact Relations}.
\newblock IOP Publishing, Bristol, UK, 2016.

\bibitem{Haus:1989:EFE}
Hermann~A. Haus and James~R. Melcher.
\newblock {\em Electromagnetic Fields and Energy}.
\newblock Pren{\-}tice-Hall, Upper Saddle River, New Jersey, 1989.

\bibitem{Ikehata:1998:SEI}
M.~Ikehata.
\newblock Size estimation of inclusion.
\newblock {\em Journal of Inverse and Ill-Posed Problems}, 6(2):127--140, 1998.

\bibitem{Johnson:1987:TDP}
David~Linton Johnson, Joel Koplik, and Roger Dashen.
\newblock Theory of dynamic permeability and tortuosity in fluid-saturated
  porous media.
\newblock {\em Journal of Fluid Mechanics}, 176(??):379--402, March 1987.

\bibitem{Kajiura:2010:SAE}
Stephen~M. Kajiura, Anthony~D. Cornett, and Kara~E. Yopak.
\newblock Sensory adaptations to the environment: Electroreceptors as a case
  study.
\newblock In Jeffrey~C. Carrier and John~A. Musick, editors, {\em Sharks and
  Their Relatives II: Biodiversity, Adaptive Physiology, and Conservation}, CRC
  Marine Biology Series, pages 393--433, Boca Raton, Florida, February 2010.
  CRC Press.

\bibitem{Kang:1997:ICP}
Hyeonbae Kang, Jin~Keun Seo, and Dongwoo Sheen.
\newblock The inverse conductivity problem with one measurement: Stability and
  estimation of size.
\newblock {\em SIAM Journal on Mathematical Analysis}, 28(6):1389--1405,
  November 1997.

\bibitem{Karlin:1966:TSA}
Samuel Karlin and William~J. Studden.
\newblock {\em Tchebycheff systems: {W}ith applications in analysis and
  statistics}.
\newblock Pure and Applied Mathematics, Vol. XV. Interscience Publishers John
  Wiley \& Sons, New York-London-Sydney, 1966.

\bibitem{Kern:2020:RCE}
Christian Kern, Owen Miller, and Graeme~W. Milton.
\newblock On the range of complex effective permittivities of isotropic
  two-phase composites and related problems.
\newblock Submitted., 2020.

\bibitem{Kirsch:2011:SOH}
Andreas Kirsch.
\newblock {\em An Introduction to the Mathematical Theory of Inverse Problems},
  volume 120 of {\em Applied Mathematical Sciences}.
\newblock Springer-Verlag, Berlin, Germany~/ Heidelberg, Germany~/ London, UK~/
  etc., second edition, 2011.

\bibitem{Kohn:1984:DCB}
Robert~V. Kohn and Michael~S. Vogelius.
\newblock Determining conductivity by boundary measurements.
\newblock {\em Communications on Pure and Applied Mathematics (New York)},
  37(3):289--298, May 1984.

\bibitem{Kohn:1987:RVM}
Robert~V. Kohn and Michael~S. Vogelius.
\newblock Relaxation of a variational method for impedance computed tomography.
\newblock {\em Communications on Pure and Applied Mathematics (New York)},
  40(6):745--777, November 1987.

\bibitem{Kolokolnikov:2015:RMS}
T.~Kolokolnikov and A.~E. Lindsay.
\newblock Recovering multiple small inclusions in a three-dimensional domain
  using a single measurement.
\newblock {\em Inverse Problems in Science and Engineering}, 23(3):377--388,
  2015.

\bibitem{Krein:1977:MMP}
M.~G. Kre\u{\i}n and A.~A. Nudel\cprime~man.
\newblock {\em The {M}arkov moment problem and extremal problems}.
\newblock American Mathematical Society, Providence, R.I., 1977.
\newblock Ideas and problems of P. L. \v{C}eby\v{s}ev and A. A. Markov and
  their further development, Translated from the Russian by D. Louvish,
  Translations of Mathematical Monographs, Vol. 50.

\bibitem{Lakes:1996:VBI}
Roderic~S. Lakes and John Quackenbush.
\newblock Viscoelastic behaviour in indium tin alloys over a wide range of
  frequency and time.
\newblock {\em Philosophical Magazine Letters}, 74(4):227--232, 1996.

\bibitem{Mattei:2016:BRL}
Ornella Mattei.
\newblock {\em On bounding the response of linear viscoelastic composites in
  the time domain: The variational approach and the analytic method}.
\newblock {Ph.D.} thesis, University of Brescia, Brescia, Italy, 2016.

\bibitem{Mattei:2017:BOP}
Ornella Mattei and Angelo Carini.
\newblock Bounds for the overall properties of composites with time-dependent
  constitutive law.
\newblock {\em European Journal of Mechanics, A, Solids}, 61:408--419,
  January\slash February 2017.

\bibitem{Mattei:2016:BRV}
Ornella Mattei and G.~W. Milton.
\newblock Bounds for the response of viscoelastic composites under antiplane
  loadings in the time domain.
\newblock In Milton \cite{Milton:2016:ETC}, pages 149--178.
\newblock See also arXiv:1602.03383 [math-ph].

\bibitem{Micchelli:1974:CCA}
C.~A. Micchelli.
\newblock Characterization of {C}hebyshev approximation by weak {M}arkoff
  systems.
\newblock {\em Computing (Arch. Elektron. Rechnen)}, 12(1):1--8, 1974.

\bibitem{Milton:1981:BCP}
Graeme~W. Milton.
\newblock Bounds on the complex permittivity of a two-component composite
  material.
\newblock {\em Journal of Applied Physics}, 52(8):5286--5293, August 1981.

\bibitem{Milton:1981:BTO}
Graeme~W. Milton.
\newblock Bounds on the transport and optical properties of a two-component
  composite material.
\newblock {\em Journal of Applied Physics}, 52(8):5294--5304, August 1981.

\bibitem{Milton:2002:TOC}
Graeme~W. Milton.
\newblock {\em The Theory of Composites}, volume~6 of {\em Cambridge Monographs
  on Applied and Computational Mathematics}.
\newblock Cambridge University Press, Cambridge, UK, 2002.
\newblock Series editors: P. G. Ciarlet, A. Iserles, Robert V. Kohn, and M. H.
  Wright.

\bibitem{Milton:2020:UPLI}
Graeme~W. Milton.
\newblock A unifying perspective on linear continuum equations prevalent in
  physics. {Part I: Canonical} forms for static and quasistatic equations.
\newblock Available as arXiv:2006.02215 [math.AP]., 2020.

\bibitem{Milton:2020:UPLII}
Graeme~W. Milton.
\newblock A unifying perspective on linear continuum equations prevalent in
  physics. {Part II: Canonical forms} for time-harmonic equations.
\newblock Available as arXiv:2006.02433 [math-ph]., 2020.

\bibitem{Milton:2020:UPLIII}
Graeme~W. Milton.
\newblock A unifying perspective on linear continuum equations prevalent in
  physics. {Part III: Canonical} forms for dynamic equations with moduli that
  may, or may not, vary with time.
\newblock Available as arXiv:2006.02432 [math-ph], 2020.

\bibitem{Milton:2020:UPLIV}
Graeme~W. Milton.
\newblock A unifying perspective on linear continuum equations prevalent in
  physics. {Part IV: Canonical} forms for equations involving higher order
  gradients.
\newblock Available as arXiv:2006.03161 [math-ph]., 2020.

\bibitem{Milton:1997:EVM}
Graeme~W. Milton and James~G. Berryman.
\newblock On the effective viscoelastic moduli of two-phase media. {II}.
  {Rigorous} bounds on the complex shear modulus in three dimensions.
\newblock {\em Proceedings of the Royal Society A: Mathematical, Physical, \&
  Engineering Sciences}, 453(1964):1849--1880, September 1997.

\bibitem{Milton:2016:ETC}
Graeme~W. {Milton (editor)}.
\newblock {\em Extending the Theory of Composites to Other Areas of Science}.
\newblock Milton--Patton Publishers, P.O. Box 581077, Salt Lake City, UT 85148,
  USA, 2016.

\bibitem{Mueller:2012:LNI}
Jennifer~L. Mueller and Samuli Siltanen.
\newblock {\em Linear and Nonlinear Inverse Problems with Practical
  Applications}.
\newblock Computational Science \& Engineering. SIAM Press, Philadelphia, 2012.

\bibitem{Prokhorov:2015:RAM}
Vasiliy~A. Prokhorov.
\newblock On rational approximation of {M}arkov functions on finite sets.
\newblock {\em J. Approx. Theory}, 191:94--117, 2015.

\bibitem{Sylvester:1993:LAI}
John Sylvester.
\newblock Linearizations of anisotropic inverse problems.
\newblock In Lassi P{\"a}iv{\"a}rinta and Erkki Somersalo, editors, {\em
  {Inverse Problems in Mathematical Physics: Proceedings of The Lapland
  Conference on Inverse Problems Held at Saariselk{\"a}, Finland, 14--20 June
  1992}}, volume 422 of {\em Lecture Notes in Physics}, pages 231--241, Berlin,
  Germany~/ Heidelberg, Germany~/ London, UK~/ etc., 1993. Springer-Verlag.

\bibitem{Sylvester:1987:GUT}
John Sylvester and Gunther Uhlmann.
\newblock A global uniqueness theorem for an inverse boundary value problem.
\newblock {\em Ann. of Math. (2)}, 125(1):153--169, 1987.

\bibitem{Tartar:2009:GTH}
Luc Tartar.
\newblock {\em The General Theory of Homogenization: a Personalized
  Introduction}, volume~7 of {\em Lecture Notes of the Unione Matematica
  Italiana}.
\newblock Springer-Verlag, Berlin, Germany~/ Heidelberg, Germany~/ London, UK~/
  etc., 2009.

\bibitem{Torquato:2001:RHM}
Salvatore Torquato.
\newblock {\em Random Heterogeneous Materials: Microstructure and Macroscopic
  Properties}, volume~16 of {\em Interdisciplinary Applied Mathematics}.
\newblock Springer-Verlag, Berlin, Germany~/ Heidelberg, Germany~/ London, UK~/
  etc., 2002.

\bibitem{Emde:2006:NVE}
G.~{von der Emde}.
\newblock Non-visual environmental imaging and object detection through active
  electrolocation in weakly electric fish.
\newblock {\em Journal of Comparative Physiology A}, 192:601--612, 2006.

\bibitem{Walsh:1932:IAR}
J.~L. Walsh.
\newblock On interpolation and approximation by rational functions with
  preassigned poles.
\newblock {\em Trans. Amer. Math. Soc.}, 34(1):22--74, 1932.

\bibitem{Walsh:1965:IAR}
J.~L. Walsh.
\newblock {\em Interpolation and approximation by rational functions in the
  complex domain}.
\newblock Fourth edition. American Mathematical Society Colloquium
  Publications, Vol. XX. American Mathematical Society, Providence, R.I., 1965.

\bibitem{Welters:2014:SLL}
Aaron~T. Welters, Yehuda Avniel, and Steven~G. Johnson.
\newblock Speed-of-light limitations in passive linear media.
\newblock {\em Physical Review A (Atomic, Molecular, and Optical Physics)},
  90(2):023847, August 2014.

\bibitem{Werner:1962:KET}
Helmut Werner.
\newblock Die konstruktive {E}rmittlung der {T}schebyscheff-{A}pproximierenden
  im {B}ereich der rationalen {F}unktionen.
\newblock {\em Arch. Rational Mech. Anal.}, 11:368--384, 1962.

\bibitem{Werner:1962:SDT}
Helmut Werner.
\newblock Ein {S}atz \"{u}ber diskrete {T}schebyscheff-{A}pproximation bei
  gebrochen linearen {F}unktionen.
\newblock {\em Numer. Math.}, 4:154--157, 1962.

\bibitem{Werner:1962:TAB}
Helmut Werner.
\newblock Tschebyscheff-{A}pproximation im {B}ereich der rationalen
  {F}unktionen bei {V}orliegen einer guten {A}usgangsn\"{a}herung.
\newblock {\em Arch. Rational Mech. Anal.}, 10:205--219, 1962.

\bibitem{Willis:1981:VRM}
John~R. Willis.
\newblock Variational and related methods for the overall properties of
  composites.
\newblock {\em Advances in Applied Mechanics}, 21:1--78, 1981.

\end{thebibliography}
\end{document}